\documentclass[12pt]{article}

\usepackage{amsmath,amssymb}
\usepackage{bm}
\usepackage{epsfig}
\usepackage{slashed}
\usepackage{graphicx}
\usepackage{caption}
\usepackage{subcaption}

\usepackage{placeins}

\begin{document}


\title{
\bf Strangeness content of the pion in the U(3) Nambu-Jona-Lasinio model}

\author{ F\'abio L. Braghin }

\author{ F\'abio L. Braghin
\\
{\normalsize Instituto de F\'\i sica, Federal University of Goias,}
\\
{\normalsize Av. Esperan\c ca, s/n,
 74690-900, Goi\^ania, GO, Brazil}
}

\maketitle 

\abstract{
The Nambu-Jona-Lasinio model is considered
with flavor-dependent coupling constants 
\\
$G_{ij} \left[ ( \bar{\psi} \lambda_i \psi ) ( \bar{\psi} \lambda_j \psi )
+   ( \bar{\psi} i\gamma_5 \lambda_i \psi ) ( \bar{\psi} i \gamma_5\lambda_j \psi )
\right]$ for 
 $i,j=0,1..N_f^2-1$,
 and $N_f=3$.
  A self consistent calculation
of quark effective masses and coupling constants is performed
making the strange quark effective mass to vary considerably.
Quantum mechanical  mixings between up, down and strange 
constituent quarks
yields a strangeness content of the light u and d quarks constituent and of the pion.
Different types 
of estimates for the strangeness contribution for the pion mass 
are provided.
Mixing type interactions, $G_{i\neq j}$, induce the light
mesons mixings and 
estimates for the $\pi^0-\eta$ and $\eta-\eta'$ mixings
are provided.
 The $\eta-\pi^0$ mixing is argued to be 
an indication of  the 
strangeness content of the pion.
}

\section{ Introduction }
 
The detailed description of hadron spectra with analytical methods
in   Quantum Chromodynamics presents many difficulties.
Usually it requires some approximate schemes  
being also possible to  resort to  effective models valid within 
a range of a variable, usually low energy, associated  to some physical scale.
In spite of the  limitations of an effective model,  when compared to first principles
calculations,
in many cases they manifest the most important degrees of freedom and
allow for a deeper understanding of Strong Interactions.
One also expects that improvements can be done 
and  eventually  may  produce a framework 
 hopefully  comparable 
to effective field theories (EFT).
This might be achieved if fundamental properties of QCD 
are taken into account by introducing the correct degrees of freedom
in a suitable and correct way.
Besides that, effective models can show very clearly the main connections between 
observable and the corresponding relevant degrees of freedom.
The quark-level Nambu-Jona-Lasinio model (NJL)   
\cite{NJL,klevansky,vogl-weise,hatsuda-etal,hua-etal}
captures some important features of quark dynamics.
It has shown to be appropriate to describe several aspects of 
hadrons dynamics, in particular the light hadron  spectra,
  whenever Dynamical Chiral Symmetry Breaking
(DChSB) plays an important role.
It provides a framework, in general,
consistent with the constituent quark model
\cite{constituent-1}.
Constituent, or dressed, quark masses are obtained with
contributions  of the  chiral condensates  that,
 added to the masses originated from the Higgs
boson, provides the correct scale of magnitude of hadrons masses
  \cite{weinberg-book,BRST-2010}.
Usually,  DChSB
is only  produced as  long as 
coupling constants are minimally strong and this imposes restrictions
in coupling constant of  the NJL model.
{A strongly interacting gluon cloud can be seen 
as  to give rise to (at least part of) the NJL coupling constant
\cite{GNJL2,coimbra-etc,kondo,PRD-2013}.}
Among 
other possible related reasons,   trace anomaly might also 
be involved in these mechanisms
\cite{traceanomaly}.
Other similar successful
calculations with contact-interactions, usually vector-current interactions
inspired in    large gluon effective mass induced interaction,
 have also been done
 \cite{SBK,craig+chinese}.
It can be expected there appears  a  (at least partial)
relation between these models because of the Fierz transformations.
Pseudo-scalar mesons mixings can be described for
broken $U_A(1)$ symmetry.
The U$_A$(1) anomaly manifests in 
Wess-Zumino-Witten terms
 and usually, for the quark-sector,
by means of the  't Hooft determinantal interaction \cite{thooft}.
This interaction is induced by the 
instanton solutions in Euclidianized Yang Mills equations and 
its role in the phenomenology has been exploited extensively in the 
mean field level 
and beyond \cite{klevansky,vogl-weise,hatsuda-etal,weise-etal,bernard-etal,eigthorder}.
It has been invesitgated extensively  in 
models exclusively with mesons degrees of freedom, for example in
\cite{eta-etap,ohta-etal,mixing-instantons}.
Mesons mixings provide a  solution of the  $\eta-\eta'$ mass   problem \cite{eta-etap}.
In the present work, we consider and investigate the consequences of 
a different effect.
Therefore 
 it will be  assumed $U_A(1)$ has been broken
although
the 't Hooft interaction will not be considered.

Light hadron spectrum is mostly  quite well described  by NJL-type models
in  agreement with the quark model.
 In addition to  valence  quarks    their  corresponding
 quark-antiquark condensates  produce important contributions.
Indeed, other   partons, besides the valence quarks,
 were shown to be needed to describe 
with precision 
hadron hadron structure  
for few examples
\cite{weinberg-book,partons-1,partons-2,EMCg1,MS+cheng-li}.
Among these partons,
strange sea  quarks might yield
a  strangeness content of the non-strange light hadron sector
by starting with  the nucleon
properties.
Results for
  electromagnetic properties of the nucleon
suggest a strangeness content (s-content)  to be of the order of $5\%$ 
\cite{exp-N-strange,magmomlatt,smmNprl,atlas-prec,DESY-exp,s-cont-latt,nature-young}.
The strange, up and down sea quark densities, however,  
were found to be nearly the same \cite{atlas-prec}.
Among the   light hadrons, pions and kaons are quasi-Goldstone 
bosons of the DChSB
and it becomes of special interest  to 
understand further their structures.
The pion mass has been shown to be due, 
more than 90$\%$, to the quark-antiquark condensate \cite{pion-mass}.
Investigations in current and future facilities, for example  
  AMBERS-CERN, EIC and JLab
 are planned to 
probe  the (parton structure of the) pion and the  kaon
and the $\eta-\eta'$ mixing (JLab) \cite{jlab-mixing}.
Earlier estimates  about the
strangeness content of the pion, 
by means of mesons loops
that reduce mesons masses \cite{meson-loops},
have shown an  extremely small contribution 
to their masses \cite{cloet-roberts}.
In the present work this subject is addressed
in the framework  of 
the NJL model.

Quark-antiquark polarization for  the NJL-model was found to provide flavor dependent
corrections
to the NJL coupling constant \cite{FLB-2021a}.
{
A microscopic origin for the NJL coupling constant, as discussed above, must 
rely only on  gluon exchange that is independent of flavor and this corresponds to 
chirally symmetric quark dynamics if quark loops are not included.
Non-degeneracy of quark masses must however manifest on quark dynamics,
and, the way it manifest  at the hadron level might involve different effects.
}
Being an effective model for QCD, it is reasonable to expect that 
all the (free) parameters of the model might be traced back to 
degrees of freedom of QCD.
At the QCD level the quark current masses are the only parameters 
containing flavor symmetry breaking.
Therefore, in  an effective model,
 contributions of the current mass differences should be expected in
all the free parameters of the effective model,
similarly to the underlying ideas  for an effective field theory  (EFT)
\cite{weinberg,gasser-leutwyler,CHPT}.
{
In \cite{PRD-2014,PLB-2016,FLB-2021a,EPJA-2018} the background field method was employed to calculate
quark-antiquark effective interactions at the one loop level.
Quark field is split in sea quarks and background quarks that might correspond
to  constituent quarks eventually.
Both the NJL-model and the Global Color Model (GCM) were considered.
The same structure obtained with  the NJL model is recovered in the 
very long wavelength limit for the GCM, in particular when
  zero momentum exchange limit is taken.
}
Preliminary perturbative estimation 
for pseudoscalar and scalar  light mesons masses
 showed that
flavor-dependent coupling constants change resulting mesons masses 
slightly less than the  flavor-dependence of quark  effective masses.
The
 sizable corrections nevertheless  improve the 
description of several observable.
In the present work, 
 a self consistent calculation of coupling constants and effective masses
will be done.
 The resulting s-content of u and d constituent quarks can be
understood in terms of the
quantum mixings  \cite{sakurai,weinberg-book}.
 The need to deal with eigenstates
of two different flavor-U(N) representations for quarks and
quark-antiquark mesons generates mixings of quarks and mesons.
Fundamental up-down-strange quark
mixings are given in terms of the Cabibbo
 angle for the Cabibbo-Kobayashi-Maskawa  (CKM) matrix
\cite{cabibbo-1,MS+cheng-li}.
 The parameterizations of quark and mesons
mixings are therefore established.

Therefore, 
in this work, the strangeness content of the u and d constituent quarks
and 
  of the neutral and charged pion masses
will be  analyzed
in the NJL model with flavor-dependent coupling constants.
{
Because we consider a polarization process with the contribution of
gluon dynamics by means of an effective gluon propagator,
there appears the need to normalize the resulting coupling constants with respect to 
the initial, standard, NJL -coupling constant.
This normalization will be different from the one adopted in  Ref. \cite{FLB-2021a}
and it will  favor a faster
 convergence of the self consistent calculation of masses and coupling constants.
As we add different components of the coupling constants
and perform a self consistent calculation for effective masses and
coupling constants
the resulting effective masses will  change considerably.
Therefore the (input)  parameters of the model either  must be redefined in a fitting 
procedure or  the coupling constants in the BSE might be
eventually dressed differently from the one in the gap equation
because of the non-renormalizability of the model.
Therefore in the BSE   considered to calculate
mesons bound state, one might need either to perform a new fit of the 
current quark masses or to truncate the equations by keeping 
a constant $G_0$ for the part of the equation that contains the quadratic divergence.
With this choice,
results become similar to the results obtained perturbatively in Ref. \cite{FLB-2021a}.
We choose the latter procedure and left the overall complete
new fit of 
parameters for another work.
}
  Different estimations  of the  strangeness contribution for 
their masses 
will be provided.
 Estimates for the $\eta-\eta'$ 
and $\pi^0-\eta$ angle mixings are also provided.
{ For these estimates, the masses of $\eta$ and $\eta'$ will not be computed,
and  the mixing angles will  be computed in a restricted way  with the mixing interactions,
$G_{i \neq j}$.
This can be done  by imposing
the corresponding meson mass differences.
For this,  the auxiliary field method  will be considered in a more general prescription than
adopted in \cite{FLB-2021a} but results are very similar.
 Light mesons mixings must be proportional to the light quark mass differences
and therefore have small amplitudes \cite{gross-etal,kaplan}.
Moreover, it will be  argued that the $\eta-\pi^0$  mixing can 
provide information about the strangeness-content of the pion, including a contribution 
for its mass.
A contribution for the pion mass 
 will be  computed by assuming  sea strange quark masses 
to be of the order of a
constituent strange quark mass in a rest frame.}
Whereas the quantum mixing is considered for the calculation
of quark effective masses and coupling constants,
the mixing type interactions, $G_{i\neq j}$ or $G_{f_1\neq f_2}$, will only be
considered for the estimation of mesons mixngs.

 The work is organized as follows.
In the next section the whole framework will be reminded
with particular attention to the definition of the coupling constants.
The logics of  the self consistent 
calculation of 
$G_{ij}$
and  quark effective masses will be emphasized.
The bound state equation (BSE), a  Bethe-Salpeter equation
at the Born level, 
for the quark-antiquark pseudoscalar mesons will  be  also briefly reminded.
{
The NJL model is a non renormalizable model intended to be valid for global properties of hadrons
at  lower energies and, as such, its calculated observables do
 depend on a chosen ultraviolet (UV) cutoff.
Moreover,  results from the NJL model are known to depend on the 
chosen regularization
scheme. 
However,  it has been found in different works 
that   the difference among the different
schemes for many observables
  are  not really large  for  light hadrons \cite{kohyama-etal,klevansky}.
In the present work the three-dimensional (Euclidean) momentum cutoff scheme
 is adopted.
}
Numerical results will be  presented in the following section
for sets of coupling constants
generated by three  different gluon propagators.
Results will be compared with a calculation for the
flavor-independent NJL  model with 
a coupling constant of reference $G_0$.
The neutral pion and kaon masses, or conversely
the charged pion and kaon masses, will be used to 
fix the set of parameters with which
further observables are also presented to assess 
the overall predictions of the 
model within the self consistent calculation.
After the self consistent calculation, that fixes the parameters,
the strange quark effective mass will be freely varied.
Nevertheless the self consistency of the up and down constituent quark
effective  masses and the coupling constants is mantained.
With this procedure one expects 
 to understand the role of the quark-antiquark
strange condensate on the up and down quark effective masses 
and pion  masses.
{The dependence of the pion decay constant with the strange quark effective mass 
is also presented.}
{Several observables, typically estimated within the NJL model,
are also calculated.
Among them,  the angle associated to the 
$\eta-\eta'$ and $\eta-\pi^0$ mixings are  provided by considering 
the flavor -dependent interactions $G_{i\neq j}$ (i,j=0,8).
{ Approximate estimations are done to reproduce the $\eta-\eta'$ and $\eta-\pi^0$ mass 
differences
 (not the complete set of neutral pseudocalar masses $\eta, \eta'$ and $\pi^0$ simultaneously)
for which one needs  $G_{08}$ and  $G_{38}$  respectively.
A strangeness-content of the pion will be  obtained from the $\pi^0-\eta$ mixing.
}
Particular values for the up and down constituent quarks and for the 
neutral pion will be  also  presented  for particular contributions of the 
strange quark condensate (or effective mass).
In the last section there is a Summary.

\section{ Masses and coupling constants: a self consistent analysis}

The generating functional of the NJL model with 
flavor dependent corrections to the coupling constants can be written as:
\begin{eqnarray} \label{g-njl-1}
Z [\eta,\bar{\eta}] = \int D_{\psi} 
exp \left[ i \int_x \; \left\{
 \bar{\psi} S_0^{-1} \psi + 
G_{ij}
[ ( \bar{\psi} \lambda_i \psi ) ( \bar{\psi} \lambda_j \psi )
+   ( \bar{\psi} i\gamma_5 \lambda_i \psi ) ( \bar{\psi} i \gamma_5\lambda_j \psi )
]
  + L_s \right\} \right],
\end{eqnarray}
where  
{$S_0^{-1}=( i \slashed{D} - m_f )$, where $\slashed{D}$ is the U(1) covariant derivative,}
$D_{\psi}= D[\psi,\bar{\psi}]$ is the functional measure,
$\int_x= \int d^4 x$,
the subscript $_{f  =  u,   d,   s}$ is used for the flavor $SU(3)$
 fundamental representation, 
 $i,j=0,...N_f^2-1$ is used for flavor indices in the adjoint representation,
being $N_f=3$ the number of flavors, and $\lambda_i$ are the flavor Gell-Mann matrices
with  $\lambda_0 = \sqrt{2/3} I$.
Quark sources are encoded in
$L_s = \bar{\eta} \psi  + \bar{\psi} \eta$.
Usually to account for the axial anomaly the 
't Hooft interaction is considered.
It is a determinant of a $N_f\times N_f$ matrix that can be written as:
 ${\cal L}_{tH} = \kappa \left( \det (\bar{\psi} P_L \psi ) 
+ \det (\bar{\psi} P_R \psi ) \right)$,
 where $P_{R/L}$ are the chirality projectors and $\kappa$ is a coupling 
constant taken as free parameter of the model.
In the $N_f=3$ model, this interaction is a 6th order quark self interaction
that has been investigated in many works 
\cite{klevansky,vogl-weise,hatsuda-etal,weise-etal,bernard-etal,eigthorder}.
{
It is interesting to note that, in this $N_f=3$ the same 6th order interaction,
except for the value of the coupling constant,
can be obtained from polarization correction for the NJL model by using 
the background field method \cite{PRD-2014,PLB-2016}.
}
In the present work all the flavor-dependent coupling constants will be 
given by $G_{ij}$ only.
{
The coupling constant 
has two components: $G_{ij} = (G_0 + \tilde{G}_{ij})$, where
$G_0$ 
is a standard NJL-coupling constant.
$G_0$ is 
 flavor independent and therefore it must be due to
 gluon dynamics.
This is a parameter of the model and a minimum critical value for it
is required to 
provide DChSB in the NJL model \cite{klevansky,vogl-weise}.
As pointed out in the Introduction there are several 
estimations of $G_0$ from  QCD degrees of freedom.}
By  means of the background field method
 in the very long wavelength limit the flavor-dependent corrections
 were found to be given by \cite{FLB-2021a}:
\begin{eqnarray} 
\label{Gij}  
\tilde{G}_{ij} &=&  
d_2 N_c (\alpha g^2)^2  \;
Tr_D  \; Tr_F
\int
 \frac{d^4 k }{ (2 \pi)^4} 
S_{0f} (k) R (k) i \gamma_5\lambda_i  S_{0f} (-k) R (-k) i \gamma_5 \lambda_j ,
\end{eqnarray}
where $Tr_D, Tr_F$ are the traces in  Dirac and flavor indices, $\alpha=4/9$, $g^2$
is the running quark-gluon coupling constant,
$d_n =  \frac{ (-1)^{n} }{2 n}$, 
$S_{0f}(k)$ is  the Fourier transform  
of  the effective  quark propagator $S_{0f}(x-y)$ 
which account for the DChSB by means of the 
quark effective mass $M_f$ or $M_f^*$ as discussed below.
In this last equation
$R(k) = 2 ( R_T (k) + R_L (k))$, where 
$R_T(k)$ and $R_L(k)$ are transversal and longitudinal 
components of an effective gluon propagator in a covariant gauge.
Other types of contributions, due to 
gauge boson dynamics and confinement, proportional to delta functions,
$\delta (p^2)$, provide smaller or vanishing contributions \cite{FLB-2021a}.
{
The corresponding Feynman diagrams of  Eq. (\ref{Gij}) are exhibitted in 
Fig. (\ref{fig:Feynman}) where the straight lines represent  quarks and
wiggly lines  with a dot  represents  non perturbative gluon propagator.
An alternative way of doing the calculation for $G_{ij}$ - eq. (\ref{Gij}) - 
would be the one-loop background field for the standard SU(3) NJL model, along 
the lines of Ref. \cite{PRD-2014}.
However in the present version, we keep track of possible contributions of the 
specific (effective) gluon propagator making possible to compare the effects
of different (effective) gluon propagators on constituent quark (or hadron) dynamics
in an effective way.

\begin{figure}[ht!]
\centering
\includegraphics[width=120mm]{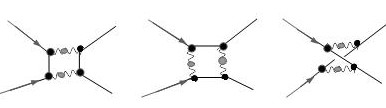}
 \caption{ \label{fig:Feynman}
\small
Feynman diagrams that correspond  to eq. (\ref{Gij}),
where the straight lines are quarks and 
wiggly lines  with a dot  represent a non perturbative (dressed) gluon propagator.
The  dots in the vertices represent the running quark-gluon coupling constant in the 
strong coupling limit.
}
\end{figure}
\FloatBarrier

}

The following important properties, 
due to CP and electromagnetic U(1) invariances, hold:
\begin{eqnarray} \label{Gij-u1}
G_{i j} = G_{j i}, \;\;\;
G_{22} = G_{11}, \;\;\; 
G_{55} = G_{44}, \;\;\;
G_{77} = G_{66}.
\end{eqnarray}
The mixing type interactions $G_{i \neq j}$ are 
proportional to quark effective mass differences
and therefore they  have considerably smaller numerical values.
{As it can be seen from eq. (\ref{Gij}) the flavor dependent coupling constants are not free parameters. The overall normalization in $G_{ij}$, however, is arbitrary 
in the same way  $G_0$ is.}
Within the usual approach for the NJL, 
scalar and pseudoscalar, $S_i, P_i$, auxiliary fields are introduced
 by means of an unit integral multiplied in the generating functional
with the corresponding shifts with quark currents that make possible
the integration of the quark field.
{
By considering the quark propagator with the  electromagnetic quark coupling,
the relations (\ref{Gij-u1}) are preserved.
This gauge invariance has the same roots of the 
description in terms of  auxiliary fields to describe electromagnetic couplings of 
charged
 mesons 
and their resulting couplings with  (background) constituent quarks 
analyzed in 
\cite{PRD-2016,EPJA-2018,JPG-2020a,JPG-2020b}.
}
In the limit of degenerate quark effective  masses, $M_u=M_d=M_s$,
the coupling constants reduce to a single constant
that, as discussed below, will be normalized to be the 
coupling constant of reference $G_0$ such that:
$G_{ij} \to G_0 \delta_{ij}$ and the standard treatment of the model can be
 done.
In this case
  the quark
effective  masses of constituent quarks are obtained with 
the contribution of the scalar-quark-antiquark condensate,
$M_f = m_f + \bar{S}_f$, where $\bar{S}_f$ are  the solutions of the
auxiliary field gap equations.
Gap equations might be found  as saddle point equations
for the scalar and pseudoscalar 
auxiliary fields $S_i, P_i$
and non trivial solutions should emerge for the (neutral) scalar
fields $S_0,S_3,S_8$.
 For the coupling constant of reference $G_0$ these equations
 can be written as:
\begin{eqnarray}
 (G1) \;\;\;\; M_f - m_f = G_0 Tr ( S_{0,f} (0)).
\end{eqnarray}
The gap equations for the model (\ref{g-njl-1}) however  receive corrections
from the flavor dependent coupling constants.
The final values of the coupling constants $G_{ij}$ therefore are
obtained from a self consistent calculation with the corrected gap equations as 
discussed below such that:
\begin{eqnarray} \label{Gij-Mf}
G_{ij} = G_{ij} ( M_u^*, M_d^*, M_s^*) .
\end{eqnarray}
In these equations for $G_{ij}$ one has the first type of 
mixing  interactions, i.e.  $G_{i\neq j}$,
that are numerically much smaller than the diagonal ones $G_{ii}$
because they depend on the differences between quark effective masses
and they will not be considered in most part of  this work.

Corrections to the coupling constant  from polarization, however,  
might produce 
spurious increasing values of the NJL coupling constant
that is a free parameters of the model.
To make possible comparisons of  numerical results 
 from different choices of the gluon propagator for $G_{ij}$
with results from  a coupling constant of reference, $G_0 = 10$GeV$^{-2}$,
 a
normalization procedure will be adopted
after the calculation of the integrals in eqs. (\ref{Gij-Mf}).
{
As discussed in the Introduction,
lately one has associated $G_0 \sim 1/M_G^2$, where $M_G$ is an effective gluon mass.
By neglecting 
further dimensionless constants, this would correspond to $M_G \sim 315$MeV, that is 
smaller than usual values obtained in lattice and SDE calculations.
However this value  for $M_G$  is   close to the value 
considered in \cite{cornwall}.
The larger value of $G_0$, when compared to usual NJL-model calculations,
favors  faster convergence of the self consistent numerical calculations.
}
 Since polarization process should produce corrections 
to an initial NJL -coupling constant, say $G_0$
the following resulting complete coupling constant should
be obtained to compute observables:
\begin{eqnarray}
G^{comp}_{ij} = \left( G_0 + \tilde{G}_{ij} \right) \bar{G}_0,
\end{eqnarray}
where $G_{ij}$ is obtained by eq. (\ref{Gij})
and $\bar{G}_0$ is a renormalization factor that 
brings the resulting value to a value of reference whenever
the symmetric limit is reached, i.e.
$G_{ij} (M^*)= G_0 \delta_{ij} =10$GeV$^{-2}$, being, in that limit,
$M^*=M^*_u=M_d^*=M_s^*$.
Besides that, the different effective gluon propagator with 
the running quark-gluon coupling constant, defined below,
have different normalizations and it becomes important to normalize
all the results by a common factor to make possible to 
understand the role of each of the variables in the set of parameters 
and effective gluon propagator.
By choosing, for example, the charged  pion mass to be a 
fitted parameter/observable
for $G_{11}^{comp} = 10$ GeV$^{-2}$, the
following normalization, written in the main text,
can be used:
\begin{eqnarray} \label{Gnorm-ij}
G_{i=j}^n \equiv G^{comp}_{i=j} = 10 \times
\frac{ G_{sym} \delta_{ij} + G_{ij} 
}{ 
 G_{sym} \delta_{ij} + G_{11} 
 }.
\end{eqnarray}
Because the coupling constants $G_{11}$ is almost equal to $G_{33}$,
it makes basically no difference to adopt neutral or charged pion mass to be a 
fitted parameter. 
In the flavor-symmetric 
  calculation for the NJL model,
  polarization effect
is also added  to the original value of $G_0$
\cite{PRD-2014,PLB-2016}.
For the mixing type interactions $G_{i\neq j}$ a similar reasoning is adopted,
 being 
that in the  flavor symmetric limit 
and in the original NJL model $G_{i \neq j}^n = 0$,
so that one can write:
\begin{eqnarray}
G_{i\neq j}^n = 10 \times \frac{G_{ij}}{G_{sym} }.
\end{eqnarray}
This normalization is compatible with 
the one for diagonal $G_{ii}$ although it is somewhat arbitrary.
This normalization (\ref{Gnorm-ij})  is different from the one considered in the 
perturbative investigation \cite{FLB-2021a} 
and the numerical results for $G_{ij}$ are somewhat similar to the ones 
presented in the perturbative case just mentioned.
Besides that,  the present normalization 
was found  to be  more  appropriated for the 
convergence of the self consistent numerical calculations.
It will be discussed that this normalization
might overestimate the role of flavor dependent interactions.
The 
  't Hooft interaction, however,  has been neglected
and results may, at the end, be reasonably close to realistic ones.

These corrections for the  NJL-coupling constants 
can re-arrange 
quark effective masses.
Restricting  to the diagonal generators, $i,j=0,3,8$,
the coupling constants  for the diagonal  flavor  singlet quark currents,
 $G_{ff}$ ($f=u,d,s$),
 can be defined
 in the following way:
\begin{eqnarray}  \label{gij-kf1f2}
G_{ij}^n (\bar{\psi} \lambda_i \psi ) 
 (\bar{\psi} \lambda_j \psi )  
= 2 \; G_{f_1f_2} (\bar{\psi} \lambda_{f_1} \psi ) 
 (\bar{\psi} \lambda_{f_2} \psi ) ,
\end{eqnarray}
where 
$\lambda_{f1}$ are three single-entry matrices obtained 
from combinations of diagonal Gell-Mann matrices with a single non zero
matrix element that are:
$\lambda_{u} = 1 e_{11}$, $\lambda_d = 1 e_{22}$ and 
$\lambda_s = 1 e_{33}$, where $e_{ii}$ is a diagonal matrix element.
The  following relations between the coupling constants $G_{ii}$ and $G_{ff}$
in the absence of the (numerically smaller) mixing-type interactions, 
$G_{i\neq j} = G_{f_1 \neq f_2} = 0$,
 are obtained:
\begin{eqnarray} \label{G-K}
 2 G_{uu}
  &=&
 2  \frac{ G_{00}^n }{3}
 + G_{33}^n + x_s \frac{G_{88}^n}{3} 
,
\nonumber
\\
2 G_{dd}
 &=& 2 \frac{ G_{00}^n}{3} 
 + G_{33}^n + x_s\frac{G_{88}^n}{3} 
,
\nonumber
\\
2 G_{ss} 
  &=& 2 \frac{ G_{00}^n }{3}
+ 4  x_s \frac{G_{88}^n}{3}  ,
\end{eqnarray}
where $x_s$ is an {\it ad hoc} parameter to control the strength of $G_{88}$.
For equal quark masses
the  flavor independent coupling constants reduce to an unique constant
$G_{ij} = G_{sym} \delta_{ij}$ and 
$G_{f_1f_2} = G_{sym} \delta_{f_1f_2}$. 
Note that for the diagonal interactions $i,j=0,3,8$ one has
 $G_{uu}=G_{dd}$.
Two cases for the strangeness content of the coupling constants will be 
considered by introducing a 
parameter $x_s$ in $G_{88}$   that provide the a contributions
from the asymmetry of strange  to up and down  quark dynamics.
Therefore an  {\it ad doc} parameter $x_s$, that controls its strength 
will be introduced by multiplying $G_{88}$
and it will be made variable to test with more details the contribution
of the strangeness in the u and d sector.
This parameter  can be
set  $x_s = 1$ at any time,  without loss of generality, in which
case one can expect to reach a physical point that describes mesons masses.
Also, it can be used to make the flavor-breaking content of $G_{88}$
  to be suppressed whenever $x_s G_{88} = 10$ GeV$^{-2}$, 
that is the value of reference for the flavor symmetric point.
The following cases will be considered  
into the equations of $G_{ff}$ as written above:
\begin{eqnarray} \label{M2M3}
(M2) \;\;\;   \; x_s G_{88} = 10 \; GeV^{-2},
\nonumber
\\
(M3) \;\;\;   \;  x_s G_{88} = G_{88} \; GeV^{-2} ,
\end{eqnarray}
In the first case, $M2$,
the role of strangeness does not take into account the flavor- asymmetry
interaction $G_{88}$ which is obtained from the eighth flavor generator 
$\lambda_8/2$. 
The second case, $M3$, is obtained with a more realistic 
account of the   strange quark
content.

The gap equations for the flavor dependent coupling constants
in the absence of  mixing-type interactions
 can be written as:
\begin{eqnarray}  \label{gap-g2}
 (G2) \;\;\;\; && {M_f^*} - m_f = G_{ff} \; Tr \; (S_{0,f} (0)),
 \end{eqnarray}
where $Tr$ includes traces in color, flavor and Dirac indices and 
momentum integral, 
and $S_{0,f}(x-y)$ is the quark  propagator
in terms of the quark effective masses $M^*_f$.

\subsection{ Mesons bound state equation
}

Pseudoscalar auxiliary fields for the  composite 
quark-antiquark  states
can describe  pseudoscalar mesons.
In particular for the case of the  pseudoscalar mesons,
the two point Green's function have pole at 
a  time-like  momentum
at zero 
tridimensional (Euclidean) momentum $\vec{P}=0$.
The NJL- model condition for the quark-antiquark pseudoscalar 
BSE can be written
 as:
\begin{eqnarray}
1 - 2 G_{ij} I_{f_1f_2}^{ij} (P_0^2=-M_{PS}^2, \vec{P}^2=0)= 0,
\end{eqnarray}
where 
\begin{eqnarray} \label{polariza-tensor}
I_{f_1f_2}^{ij} (P_0, \vec{P})
=
i Tr_{D,F,C}
\int \frac{d^4 k}{(2\pi)^4} 
\lambda_i i\gamma_5 S_{0,f_1} (k+ P/2) \lambda_j i \gamma_5 S_{0, f_2}(k- P/2),
\end{eqnarray}
where the different flavor indices for the case of the pion bound states
are the following:
$\pi^0$ with $i,j=3$ and $f_1, f_2 = u, d$ 
and 
$\pi^\pm$ with $i,j=1,2$ and $f_1, f_2 = u, d$.
Kaons and some of the scalar mesons were discussed in \cite{FLB-2021a}.
{ After the traces in Dirac, color and flavor indices  have been calculated
the equation is Wick rotated to the Euclidean momentum space-time
and the condition for obtaining the mesons masses become
$P_0^2 = - M_{PS}^2$, where 
$M_{PS}$ is the mass of the pseudoscalar meson.
}

{
The gap equations can be used to eliminate the quadratic
 divergence of $I_{f_1f_2}^{ij}$.}
In particular for the case of the  pions and kaons,
 one can write the following reduced equation:
\begin{eqnarray} \label{BSE-Gff-ii}
 (M_{PS}^2 -  ({M_{f_1}^*}  - {M_{f_2}^*}) ^2 ) G_{ij} 
I_2^{f_1f_2} 
 &=&
\frac{G_{ij}}{2}  \left( 
\frac{m_{f_1}}{ \bar{G}_{f_1f_1} M_{f_1}^* }
+ \frac{m_{f_2}}{ \bar{G}_{f_2f_2} M_{f_2}^* }
\right)
+ 1
\nonumber
\\
&-&  \frac{1}{2}
\left( \frac{G_{ij}}{\bar{G}_{f_1f_1} } + 
\frac{G_{ij}}{\bar{G}_{f_2f_2} } \right) ,
\end{eqnarray}
{where $\bar{G}_{f_1f_2}$ is the  normalized coupling constant
from the gap equations. To
cope with the need of different renormalizations for the gap eqs. and BSE,
the quark condensate from the gap eqs. will be renormalized by
$\bar{G}_{f_1f_2}/G_0$  corresponding to the choice:
$\bar{G}_{f_1f_2} \to  G_0$ in these BSE.
This guarantees the correct order of magnitude of the resulting 
neutral and charged pions and kaons masses.
}
In this equation there as a logarithmic divergent integral
given by:
\begin{eqnarray}  \label{int-bse}
I_2^{f_1f_2}  &=& 4  N_c \int \frac{d^3 k}{(2\pi)^3} 
\frac{ (E_{f_1} + E_{f_2} )}{ E_{f_1} E_{f_2} (M_{PS}^2 - (E_{f_1}+E_{f_2})^2) },
\end{eqnarray}
where  $E_{f} = \sqrt{ \vec{k}^2 + {M_f^*}^2 }$
  in Euclidean momentum space.
These integrals are  solved 
with  the same 3-dim  cutoff $\Lambda$
of the gap equations,

\section{ Numerical results}

Flavor dependent coupling constants were  calculated 
by considering three different effective gluon propagators
each of the two different sets
 {of current quark masses and UV cutoff: $S$ and 
$V$.
These effective gluon propagators will be labeled by:
$\alpha=2, 5$ and  $6$.
They incorporate the quark-gluon running coupling constant $g^2$
as shown below.
The first effective  gluon propagator ($2$)
}
  is  a transversal one
 extracted from 
Schwinger Dyson equations calculations  
\cite{gluon-prop-sde,SD-rainbow}.
 It  can be written as:
\begin{eqnarray} \label{gluon-prop-sde}
(S_2, V_2) : \;\;\;  D_{2} (k)  = g^2 R_T (k)  &=& 
\frac{8  \pi^2}{\omega^4} De^{-k^2/\omega^2}
+ \frac{8 \pi^2 \gamma_m E(k^2)}{ \ln
 \left[ \tau + ( 1 + k^2/\Lambda^2_{QCD} )^2 
\right]}
,
\end{eqnarray}
where
$\gamma_m=12/(33-2N_f)$, $N_f=4$, $\Lambda_{QCD}=0.234$GeV,
$\tau=e^2-1$, $E(k^2)=[ 1- exp(-k^2/[4m_t^2])/k^2$, $m_t=0.5 GeV$,
$D= 0.55^3/\omega$ (GeV$^2$) and 
$\omega = 0.5$GeV.

The second type of effective  gluon propagator 
 is based in a longitudinal  effective confining parameterization
\cite{cornwall}
that can be  written  as:
\begin{eqnarray} \label{cornwall}
(S_{\alpha=5,6} , V_{\alpha=5,6} ) : \;\;\; D_{\alpha=5,6} (k)  = g^2 R_{L,\alpha} (k) &=& 
\frac{ K_F }{(k^2+ M_\alpha^2)^2} ,
\end{eqnarray} 
where 
$K_F = (0.5 \sqrt{2} \pi)^2/0.6$ GeV$^2$, as considered in previous works
\cite{PRD-2019} to describe several mesons-constituent quark 
effective coupling constants and form factors.
However  different  effective gluon masses can be tested \cite{oliveira-etal}
such as a constant one: 
($ M_5 = 0.8$ GeV) or  a running effective mass given by:
$M_6  = M_6 (k^2)=  \frac{0.5}{1 +  k^2/\omega_6^2 }$GeV
for $\omega_6=1$GeV.

The 
 sets of (free) parameters, that reproduce
the neutral pion and kaon masses after the self consistent calculation, 
are given
 in Table (\ref{table:masses}): current quark masses 
$m_u, m_d$, $m_s$ and the ultraviolet (UV) cutoff $\Lambda$.
In this Table
the  resulting effective masses from the gap equation $(G1)$, 
for $G_0$, are also presented.
It is important to emphasize that the only role of the
effective  gluon
propagator is to produce numerical results for 
$G_{ij}$.
The sets of parameters $S$ and $V$ yield   the same overall behavior
of results when $M^*_s$ is varied,
therefore figures will be exhibited  only for   $S$.
Having obtained these fittings from the self consistent calculation,
 for
different sets of $G_{ij}$,
 kaons are neglected  and the  investigation  of the 
contribution of the strangeness is done.
For this,   the free-variation 
of the effective mass, ${M_s^*}$,  will be done, by keeping  the 
 effective masses of  the up and down quarks
calculated self consistently.
The values of the  mesons masses 
at the physical point, obtained  from the sets of parameters of 
Table (\ref{table:masses}), will be shown below
in 
Table (\ref{table:obsv}).
The strange  quark current  mass, $m_s$,
 is not really  relevant for the pion observables,
but  it helps to define the  physical point, 
where  kaon masses are obtained,
and to  keep track  of the value of the
strange quark-antiquark condensate.
{ 
For the chosen three-dim regularization scheme 
 the resulting values of $\Lambda$ are not considerably larger than the  resulting effective masses.
Note however that 
the cutoff is used only  for the three-momentum component,
contrarily to the other regularization schemes for which the cutoff applies for
the four-momenta \cite{klevansky,kohyama-etal}.
Therefore it is natural to expect a lower value for the cutoff in the three-dim regularization scheme.
}

\begin{table}[ht]
\caption{
\small Sets of parameters:
  Lagrangian  quark masses, ultraviolet  cutoff
and the  quark effective masses obtained from 
an initial  NJL-gap equation ($G1$) for $G_0=10$GeV$^{-2}$.
} 
\centering  
\begin{tabular}{c | c c c c  | c c c }  
\hline\hline  
set  of  & 
$m_u$ & $m_d$ & $m_s$ &
$\Lambda$  & $M_u$ &   $M_d$ &  $M_s$
\\
parameters &  MeV  &       MeV     &    MeV  & MeV   &       MeV     &    MeV  & MeV
 \\
\hline 
 $S$  &
3 &  7 &  133 &  680  &  405    &  415  & 612
\\ [0.5ex]
$V$  & 
3 & 7 & 133  &  685  &  422   &  431   & 625
\\[1ex] 
\hline 
\end{tabular}
\label{table:masses} 
\end{table}
\FloatBarrier

\subsection{ Up and down quark 
effective masses and quark-antiquark coupling constants dependencies on $M_s^*$}

In figures (\ref{fig:Guu})  and (\ref{fig:Gss})    
results for the self consistent calculation for 
the up and strange flavor-dependent coupling constants,
$G_{uu}$ and $G_{ss}$, are presented
   as functions of the (freely-varied) strange quark effective mass $M^*_s$
for the sets $S_2, S_5$ and $S_6$
and for the parameterizations $M2$ and $M3$.
All the resulting $G_{ff}$ are normalized in the flavor symmetric point
according to eq. (\ref{Gnorm-ij}), i.e.
$G_{ff} ( M^*= m^*) = 10$GeV$^{-2} $ 
 when $M_u^*=M_d^* =M_s^* = m^*$.
Results are sounder physically for $M_s^* \geq m_s \simeq 0.133$ GeV,
below this value, $M_s^*$   represents nearly 
 a variable strange quark current mass in the absence of self-consistency.
Result for the down quark is the same as $G_{uu}$,
according to eqs. (\ref{G-K}).
The different coupling constants $G_{ij}$ obtained for the different effective gluon propagators
($S_2, S_5$ and $S_6$)
may 
produce quite different numerical results (in particular for 
$M3$) although the overall behavior is 
basically the same.
The behavior for 
small and large strange quark effective mass limits are quite different 
depending on the set $S_2, S_5$ or $S_6$.
For lower values of $M_s^*$,  the sets $S_2$ and 
 $S_6$ that have larger variations than $S_5$.
For larger strange quark effective  masses
the quantities $G_{ff}$ become smaller and tend to reach finite definite values
at $M_s^* \to \infty$
that tend to be nearly independent of the gluon propagator.
Note that   the  strange quark effective mass that produces correct
values for the kaon masses is around $0.550-0.580$ GeV (for $M3$), shown 
in the Table \ref{table:obsv-shifts}) below, for all the sets $S_2,S_5,S_6$
and for $V_2, V_5, V_6$.
These large variations of $G_{ff}$ with $M^*_s$ 
may be  indication that the (re)normalization prescription 
in eq.  (\ref{Gnorm-ij}) overestimates the role of the 
flavor symmetry breaking.

\begin{figure}[ht!]
\centering
\includegraphics[width=130mm]{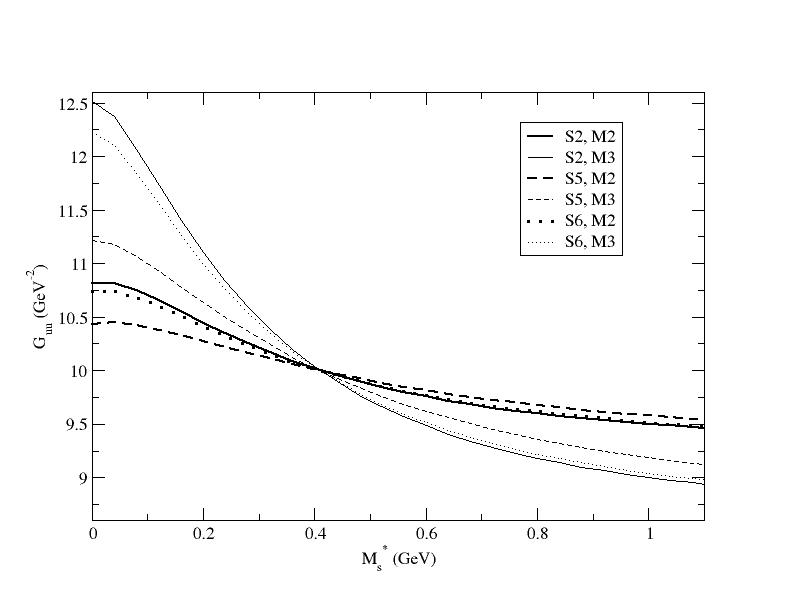}
 \caption{ \label{fig:Guu}
\small
The up quark coupling constant $G_{uu}$, eq. (\ref{G-K}),
  as a function
 of ${M_s^*}$, arbitrarily varied.
Up and down quark masses are 
obtained self consistently from their gap equations (G2).
}
\end{figure}
\FloatBarrier

\begin{figure}[ht!]
\centering
\includegraphics[width=130mm]{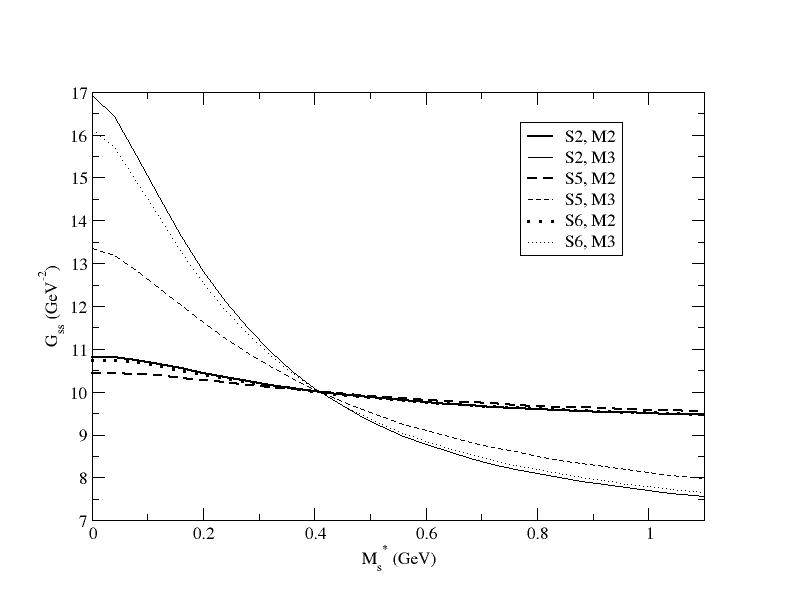}
 \caption{ \label{fig:Gss}
\small
The strange quark coupling constant $G_{ss}$, eq. (\ref{G-K}),
  as a function
 of ${M_s^*}$, arbitrarily varied.
Up and down quark effective masses are 
obtained self consistently from their gap equations (G2).
}
\end{figure}
\FloatBarrier

In figure (\ref{fig:Mu})  
the
 up quark effective mass, as self consistent solution to the gap $(G2)$,
 eq. (\ref{gap-g2}),
is presented
as a function  of the strange quark effective mass ${M_s^*}$
that is made to vary freely.
Again, the figure has a clear meaning for $M^*_s > m_s$.
The point in which all the 
cases  coincide is the symmetric point
$G_{ij} = G_{sym} \delta_{ij}$ due to the normalization adopted.

The self consistency is implemented for both cases $M2$ and $M3$ that 
controls the strangeness dependence of $G_{88}$.
The "physical value " of $M^*_s$, i.e.,
the value  that reproduces the correct kaons masses, being 
solution of the gap equation $G2$,  is around
$0.500-0.600$GeV depending on set $S2,S5, S6$, 
as  presented in  Table  (\ref{table:obsv}).
This self consistence procedure lowers 
the 
  values of the   quark effective masses.
It is  interesting to note that both limits, zero and very large 
strange quark  effective mass,
might be, in different ways,  somehow associated to
 absence of strangeness in the up and down quark dynamics.

\begin{figure}[ht!]
\centering
\includegraphics[width=120mm]{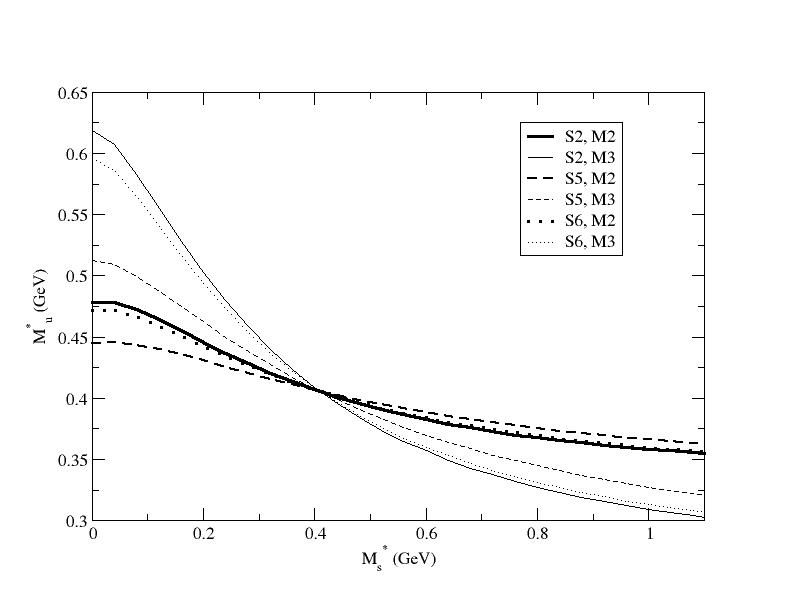}
 \caption{
\small
Self consistent solutions for the up quark effective mass, $M_u^*$,
obtained from the  gap equation $(G2)$ (\ref{gap-g2}),
 as a function of $M_s^*$ that  is arbitrarily varied.
}
 \label{fig:Mu}
\end{figure}
\FloatBarrier

In figure (\ref{fig:DifMdMu}) the difference between
 the self consistent solutions of 
 down and up quark effective masses, $M_d^* - M_u^*$,
is presented as a function of the strange quark effective mass $M^*_s$ 
that is 
freely varied.
Again it is important to stress that  the quantity  $M_s^* \geq m_s$   corresponds
to varying the strange chiral condensate arbitrarily.
It can be seen that the 
parameterizations $M3$ (smaller symbols) yield much larger  variation of 
$M_d^* - M_u^*$
mainly for the case
of smaller strange quark masses.
The behavior with $M_s^*$
 is the opposite
of the individual quark effective  masses  
$M_u^*, M_d^*$.
The difference in the results between the sets $S2,S5,S6$ 
(i.e.
effective gluon propagator)  reaches around  only 1 MeV either 
for $M2$ or $M3$, that is of the order of $10\%$ of the 
 effective mass difference.

\begin{figure}[ht!]
\centering
\includegraphics[width=120mm]{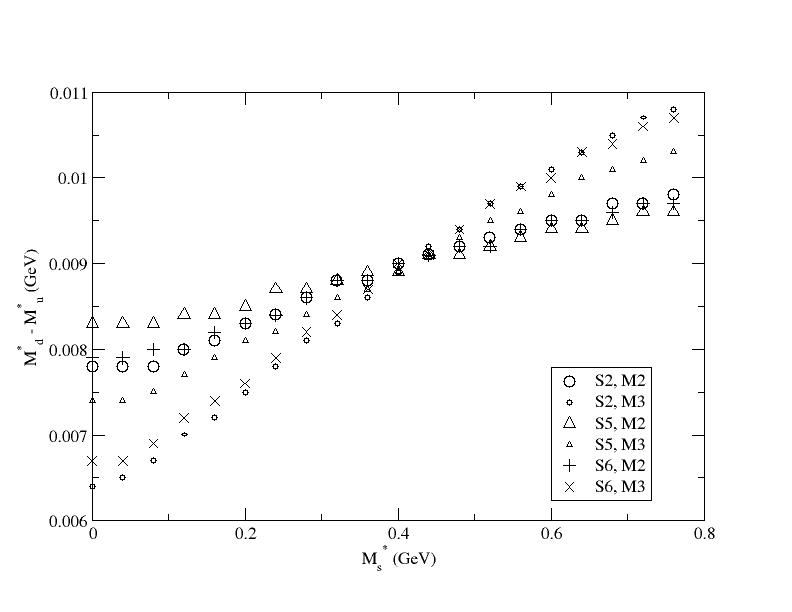}
 \caption{ \label{fig:DifMdMu}
\small
Self consistent  values for the down and up quarks effective mass difference ($M_d^* - M_u^*$)
as a function of $M_s^*$, arbitrarily varied,
both of them obtained from the  gap equations $(G2)$
eqs.  (\ref{gap-g2}).
}
\end{figure}
\FloatBarrier

\subsection{ Pion mass dependence on $M_s^*$}

In figure (\ref{fig:Mpi0})
the 
neutral pion mass as a function of the strange quark effective mass 
is exhibited for 
the self consistent  values of  $M_u^*, M_d^*$
and coupling constants
and for $M_s^*$ freely varying.
Because of  the normalization adopted,
the point in which all the 
cases  coincide is the symmetric point for which
$G_{ij} = G_{sym} \delta_{ij}$ .
The same sets of parameters $S2, S5$ and $S6$ were considered 
for the two parameterizations $M2$ and $M3$.
Both limits of  strange quark mass  going to zero and going to infinite 
are well defined, although some points were left out of the figure 
to emphasize the behavior 
for $M^*_s > m_s$, i.e. for the strange quark condensate.
It is seen that the variation of the pion mass with
the strange quark effective mass is larger for smaller
strange quark masses, i.e. smaller or vanishing strange 
quark condensate.

\begin{figure}[ht!]
\centering
\includegraphics[width=120mm]{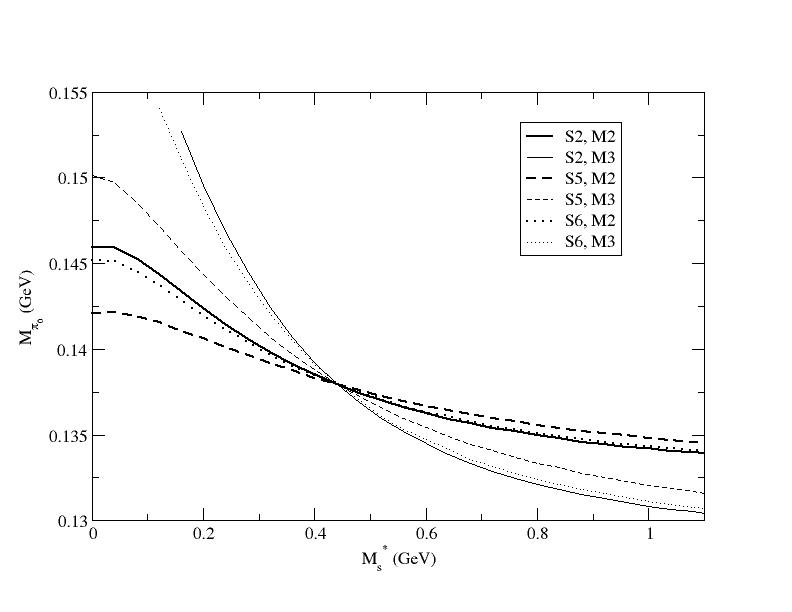}
 \caption{ 
\small
$M_{\pi_0}$ as a function of $M_s^*$, arbitrarily varied.
All the other parameters $M_u^*,M_d^*$
 and coupling constants are obtained self consistently.
}
\label{fig:Mpi0}
\end{figure}
\FloatBarrier

 Finally, in the  figure (\ref{fig:DiffMpic0})
the mass difference of charged and neutral pions, 
 $\Delta M_\pi = M_{\pi^\pm} - M_{\pi^0}$,
is exhibited as a function of the strange quark effective mass,
arbitrarily varied.
The neutral and charge pion mass difference is  known to have a 
larger contribution from electromagnetic interactions and 
only  a  small counterpart from strong interactions.
The value obtained  in quite in agreement with 
known values \cite{parada-etal,gasser-leutwyler-82,donoghue,pion-kaon-em,gasser-leutwyler-1985}.
Whereas parameterizations $M3$, smaller symbols,
provide small values of  $\Delta M_\pi$ 
for smaller $M_s^*$,
for  large strange quark masses
parameterizations $M3$ tends however to produce an increase considerably larger 
than $M2$. 
$M2$ ($M3$) makes
the mass difference  to reach a maximum value close to $0.15-0.20$MeV
($0.15-0.27$MeV) for $M_s^* > 0.8$GeV.

\begin{figure}[ht!]
\centering
\includegraphics[width=120mm]{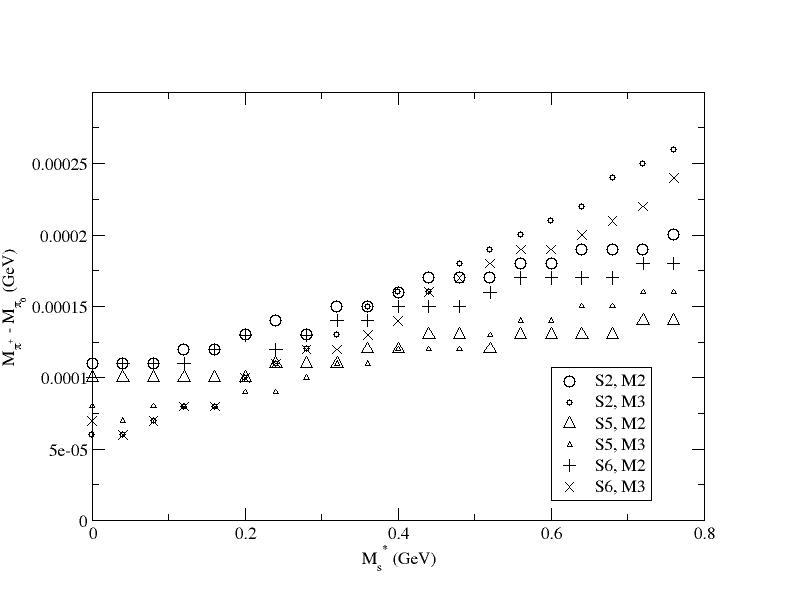}
 \caption{ 
\small
$\Delta M_\pi = M_{\pi^\pm} - M_{\pi^0}$
 as a function of $M_s^*$, arbitrarily varied.
All the other parameters $M_u^*,M_d^*$
and coupling constants
 are obtained self consistently.
}
\label{fig:DiffMpic0}
\end{figure}
\FloatBarrier

\subsection{ $\eta-\eta'$ and $\eta-\pi^0$ mixings}

The  
pseudoscalar mesons mixings will be discussed next.
 For this, the explicit
mixing interaction $G_{i\neq j}$ will be considered.
 The $\eta-\eta'$ mass difference will be 
obtained
   by means of the flavor-dependent coupling constants
 $G_{08}$.
The  auxiliary fields can be introduced by means of 
functional delta functions 
in  the generating functional \cite{alkofer-etal,osipov-etal}, for 
the case of the pseudoscalar fields one can write:
\begin{eqnarray} \label{deltafuncP}
1 \;  = \; \int   D [ P_i ] \; \delta \left( P_i - G_{ik}  j_{ps}^k \right),
\end{eqnarray}
where $j_{ps}^k = \bar{\psi} \lambda^k i \gamma_5 \psi$,
 $i,k=0,3,8$ provides the needed components to describe
the mesons $\eta, \eta'$ and $\pi_0$, and 
where the fields  dimensions  are properly taken into account.
This method neglects possible non factorizations \cite{PRD-2015}
which, nevertheless,  can be expected to be  small.
This definition reduces to 
 the usual auxiliary fields when mixing
interactions are neglected.
For  the mixing-type interactions $G_{i \neq j}$, for $i,j=0,3,8$,
one can neglect the smaller one, $G_{03}$.
 Next the corresponding quark-antiquark states 
masses and mixings can be  written in 
the adjoint representation, $M_{ii}^2 P_i^2$, in a diagonalized form.
In  general, the following quadratic terms
from the pseudoscalar auxiliary fields with the mixing interactions  $G_{ij}$
can be written:
\begin{eqnarray} \label{P0P8-mix}
{\cal L}_{mix} &=&
- \frac{M_{88}^2}{2} P_8^2  
- \frac{M_{00}^2}{2} P_0^2 + 
2  G_{08}^n \bar{G}_{08}  P_0 P_8 
+ {\cal O} (P_3, P_3^2) ...
\end{eqnarray}
where $M_{ii}^2$ include the contributions from $G_{i=j}$ derived above,
and 
\begin{eqnarray} \label{mix-08}
\bar{G}_{08} = \frac{2
 }{  G_{00}^n
\left( G_{88}^n - \frac{ {G_{08}^n}^2}{G_{00}^n } \right)
},
\end{eqnarray}
where 
the mixing terms $G_{i\neq j}$  are exclusively obtained from the 
one-loop polarization.
As seen in eq. (\ref{mix-08})
 and  the flavor dependent coupling constants $G_{ij} \propto N_c$,
 as $N_c\to \infty$ one has degenerate  $\eta$ and $\eta'$
\cite{eta-etap}.
 
The change of basis from the singlet  flavor states  basis $|\bar{q} q>$ (q=u,d,s),
 or correspondingly
$P_3,P_8,P_0$,
to the mass eigenstates $\pi^0,\eta, \eta'$  can be written as 
\cite{feldman-etal-kroll,pi0-eta-etap}:
\[
 \left( \begin{array}{c  } \label{matrix}
\pi^0
\\
\eta
\\
\eta'
   \end{array} \right)
= 
 \left( \begin{array}{c c c }
1 &   \sqrt{\frac{2}{3} } 
\epsilon_1 - \epsilon_2 \sin(\theta_{ps})
 &  \frac{ \epsilon_1}{\sqrt{3}}   + \epsilon_2 \cos(\theta_{ps})
\\
-\epsilon_2-\epsilon_1 \left(\frac{\cos(\theta_{ps})}{\sqrt{3}} -\sqrt{\frac{2}{3}}\sin(\theta_{ps})
 \right)  & - \sin (\theta_{ps})
 &   \cos (\theta_{ps})
\\
-\epsilon_1 \left(\sqrt{\frac{2}{3}} \cos(\theta_{ps}) + \frac{\sin(\theta_{ps})}{\sqrt{3}}
\right) &
\cos(\theta_{ps}) & 
\sin (\theta_{ps})
   \end{array} \right)
 \left( \begin{array}{c  }
P_3
\\
P_0
\\
P_8
   \end{array} \right).
\]
where
the parameters
 $\epsilon_1,\epsilon_2$ are mixing parameters from the Standard model.
The two sectors with larger mixings will be addressed:
the $\eta-\eta'$ mixing, that reduces to a rotation between
$P_8$ and $P_0$, and the
$\eta-\pi^0$ mixing.
By performing the usual rotation to
mass eigenstates $\eta, \eta'$, according to the 
convention from   \cite{PDG},
it can be written:
\begin{eqnarray} \label{mix-rot}
| {\eta} > &=& 
\cos \theta_{ps} | P_8 > - \sin \theta_{ps} | P_0 >,
\nonumber
\\
| {\eta}' > &=& 
\sin \theta_{ps} | P_8 > + \cos \theta_{ps} | P_0 >.
\end{eqnarray}
 Although one   needs two parameters/angles to describe both masses, $\eta, \eta'$
 \cite{mix-latt},
in this work only the mass difference will be calculated. 
It is directly due to the 
mixing-type  interaction $G_{08}$.
 By calculating ${\cal L}_{mix}$ in this mass eigenstates basis,
  and comparing to the above $0-8$ mixing,
the following $\eta-\eta'$ mixing angle is obtained:
\begin{eqnarray} \label{thetaps08}
  \theta_{ps}  = \frac{1}{2} \arcsin \left( 
\frac{ 4  G_{0 8}^n \bar{G}_{08}  }{   (M_\eta^2 - M_{\eta'}^2 )  }  \right).
\end{eqnarray}
This equation provides numerical results similar to the equation
used in \cite{FLB-2021a} being however more complete.

Besides the (leading) mixing that describes $\eta-\eta'$ puzzle, the neutral pion also mixes
with both $\eta,\eta'$ being the coupling to $\eta$
 much larger than  the coupling to $\eta'$ \cite{feldman-etal-kroll,pi0-eta-etap}.
 The following rotation to define the physical
meson fields
will be considered:
 \begin{eqnarray} \label{mix-rot-p0-p3}
| {\eta} > &=& 
\left(
-\epsilon_2-\epsilon_1 \left(\frac{\cos(\theta_{ps})}{\sqrt{3}} -\sqrt{\frac{2}{3}}\sin(\theta_{ps})
 \right)  
\right)
 | P_3 >  + 
\cos (\theta_{ps}) | P_8 >,
\nonumber
 \\
|\pi_0 > &=& | P_3 > +  
\left(  \frac{\epsilon_1}{\sqrt{3}} + \epsilon_2
\cos(\theta_{ps}) \right) | P_8 >.
\end{eqnarray}
where $\epsilon_2 \simeq \epsilon$ is  the 
usual parameter for this mixing
when neglecting non leading mixing \cite{feldman-etal-kroll,pi0-eta-etap,PDG}.
The resulting mixing parameter can be written as:
\begin{eqnarray}
  \epsilon_2   = -  \frac{1}{2} \arcsin \left( 
\frac{ 4  G_{38}^n \bar{G}_{38}  }{   (M_\eta^2 - M_{\pi^0}^2 ) 
\sin (\theta_{ps}) }\right),
\end{eqnarray}
where 
\begin{eqnarray} \label{mix-03}
\bar{G}_{38} = \frac{2
 }{  G_{88}^n
\left( G_{33}^n - \frac{ {G_{38}^n}^2}{G_{88}^n } \right)
}.
\end{eqnarray}
This eq. is analogous to the equations  for the $\eta-\eta'$ channel above.
 The predictions for $\epsilon_2$ will be shown below
being consistent with 
the estimation \cite{mixing-exp-1}:
$<\pi^0| H | \eta > \propto (m_u-m_d)$.

\subsection{ Other observables}

In this section  some of the observables calculated 
with the resulting quark and mesons masses and coupling constants
are described   and their values
are displayed in Table II.

The quark-antiquark scalar condensate, chiral condensate,
is defined as:
\begin{eqnarray}
< (\bar{q} q)_f > \equiv - Tr  \left( S_{0,f} (k) \right),
\end{eqnarray}
and therefore is directly calculated by means of the solutions for the gap equations
for the three flavors.
Values in the Table correspond to the final self consistent solution
of the (G2), i.e. eq. (\ref{gap-g2}).
These values improve initial estimation 
when calculated for  $G_0$ and $M_f$. 

The quark-meson, pion or kaon, coupling constants, or correspondingly
the normalization of the field,  obtained from the residue of the 
pole of the vertex can be written as \cite{klevansky,vogl-weise}:
\begin{eqnarray}
G_{qq PS} =  \left( \frac{ \partial \Pi_{ij} (P^2)
 }{\partial P_0^2 } \right)^{-2}_{P_0^2\equiv -M_{ps}^2},
\end{eqnarray}
where the values were calculated at the
physical mesons masses  $P_0^2= - M_{ps}^2$, and 
the polarization tensor was written in eq. (\ref{polariza-tensor}), 
$\Pi_{ij} (P^2)  = I_{f_1f_2}^{ij} (P_0^2, \vec{P}^2)$.

The  weak decay constant
of  charged mesons, pion and kaon, $F_{ps}=F_\pi, F_{K}$,
can be calculated as \cite{klevansky,vogl-weise}:
\begin{eqnarray}  \label{fps}
F_{ps} =  \frac{N_c  \; G_{qq PS}}{4}\;
\left.  \int \frac{ d^4 q}{(2 \pi)^4}
 Tr_{F,D} \left( \gamma_\mu \gamma_5 
\lambda_i \; S_{f_1} (q + P/2) \lambda_j S_{f_2} (q - P/2) \right) 
\right|_{P^2_0 = - M_{ps}^2},
\end{eqnarray}
where $f_1,f_2$ correspond to the quark/antiquark of the  meson 
and $i,j$ are the associated flavor indices as discussed for eq. (\ref{polariza-tensor}).

In Table (\ref{table:obsv}) several observables 
calculated for the sets of parameters shown above 
($S$ and $V$ for the three different gluon propagators
2,5 and 6 and $G_0$) are exhibited.
The neutral pion and kaon masses were fitted
to values  close to the 
experimental value, $M_{\pi^0} = 135$ MeV
and $M_{K^0} = 498$ MeV. 
For the case of the pion, there are two estimates, one for 
the set  $M2$ and the other from $M3$.
The flavor-dependent coupling constants, however,
tends to lower the mesons masses with respect to their values 
calculated with $G_0$.
All the other observables,
 are obtained from the more complete calculation with  $M3$.
Self consistency, 
and a more complete  account of 
strangeness in the coupling constants by means of $M3$,
lead to 
 lower values of the mesons  masses and quark effective masses 
and the need of a larger value of the UV cutoff.
The initial fit of 
the neutral pion mass, for the value of reference  for the  coupling constant
$G_0=10$ GeV$^{-2}$, was found to be $M_{\pi^0}=136.4-137.1$ MeV.
The final value with the flavor-dependent coupling constants and strange effective 
mass close to value that  reproduce a physical point, for $M3$ goes to
 $M_{\pi^0} = 133-135$ GeV,
whereas for $M2$ it goes to $M_{\pi^0} = 135.0-136.3$ GeV.
Although the sets $M2$ pin down the correct (expected) value
the idea is to show the effect of the mixing by comparing 
with the more complete result from $M3$.
The estimation for the  charged pion and kaon masses, 
$M_{\pi^{\pm}}, M_{K^\pm}$, 
are in quite good agreement with experimental or expected values.
Note that, electromagnetic effects are not taken into account,
and the expected mass differences due to strong interactions
effects have opposite signs and they are 
respectively $M_{\pi^\pm} - M_{\pi^0}  \simeq 0.1$ MeV
and 
 $M_{K^\pm} - M_{K^0}  \simeq - 5.3$ MeV 
\cite{gasser-leutwyler-82,donoghue,pion-kaon-em}.
The kaon masses are exhibited for the sake of completeness 
to show the entire set of observables used to fit parameters
in the self consistent part of the calculation
and the prediction for the neutral-charged meson mass difference.

The values of the charged pion and kaon decay constants $F_\pi, F_K$
are not far from the experimental/expected
 values (e.v.) although the choice made for the fitting 
yielded a much better value for the  kaon decay constant than $F_\pi$,
differently from usual results in the literature, see e.g. in \cite{SBK} and 
references therein. These results show an improvement with 
respect to the standard NJL model treatment.
The up, down  and strange quarks condensates, 
$<\bar{u}u>, <\bar{d} d>$
and $ < \bar{s}s>$,
and the pion (kaon)-quark coupling constants $G_{\pi qq}$ ($G_{K qq}$)
are also presented.

The last observable shown in the Table are the mixing angles
that were precisely the only quantities 
calculated with the mixing type interactions.
 The pseudoscalar  mesons
mixing angle $\theta_{ps}$ and the 
$\pi^0-\eta$ mixing angle $\epsilon_2$.
 The following masses were considered for calculating the differences:
 $M_\eta=548$MeV, 
$M_{\eta'}=958$ MeV  and $M_\pi  135$ MeV \cite{PDG}.
These values still   eventually may change further by 
other  effects, mainly  for a complete self consistent calculation
 for all types of mixings and if one considers a 't Hooft type interaction
that re-arranges the mixing-type interactions. 
{The $\pi^0-\eta$ mixing parameter, $\epsilon_2$, is   exhibited
for two situations: (I) $\theta^{ps}=15^\circ$ \cite{PDG} and 
(II) $\theta^{ps}$ as calculated from eq. (\ref{thetaps08}). 
Whereas the estimate (I) is very close to other values found in the literature,
the estimates (II) are considerably larger.

Although the values of all  the observables are not as close as they could be
to the experimental or expected values they are somewhat improved with 
respect to the flavor independent calculation which is represented in 
the Table by the column for the set of parameters with $G_0$.
It is important to stress that, the coupling constants of reference is slightly
larger than usual values
and  this makes the 
values of the condensates to be larger than they should.
Large values of $G_0$, however,  favor the convergence of the self consistent solutions for 
masses and coupling constants.
A criterium for analyzing the s-content of  the pion 
is  the probability of finding a sea s-quark  in it,
denoted by  Pr-s-content $\pi$.
It can be approximately defined by means of the change in the
pion normalization,
$Z_\pi = G_{qq\pi}^2$, with respect to the 
calculation with $G_0$.
The most relevant reason for  this change in the normalization is the 
variation of the strange quark effective mass (condensate).
However, due to 
the self consistency of the problem, there might have other
much smaller contributions due to up and down quarks.
These values are considerably  larger than the estimation
from ref. \cite{cloet-roberts} based on meson loops
which have shown a probability of finding a s-quark in a up or down dressed quark
to be of the order of $2-4\%$.

Finally, in the last two last lines 
the reduced chi-square for each of the set of parameters
for two different situations are presented always for calculation with $M3$.
Firstly  the chiral condensates 
are taken into account, 
resulting in  ten observables being  two fitted observables,
and secondly 
if the condensates  are neglected, 
it provides seven observables with  two fitted observables.
The e.v. value for pseudoscalar mixing angle and quark-antiquark condensates
 were  taken to be the average value of those shown in the last column of the Table.
These two different estimations of the chi-square were done because,
although the resulting values of the chiral condensates are improved with 
respect to the flavor-independent calculation,
the deviations of their (corrected) values are still large 
with respect to the e.v. and this makes $\chi^2_{red}$ to be 
very large.
Another source of increase of the chi-squared 
are the values  of $F_{\pi}$.
Although the pion decay constant was not really in agreement with 
experimental value, the important point is that the 
relative results  $F_{K}-F_\pi$ is slightly improved with respect to standard NJL,
that corresponds to  the  set of parameters for  $G_0$.
As discussed above, the relatively large value of $G_0$
may responsible for these discrepancies.
Also, further terms in the gap equations, eventually
due to higher order interactions or vector 
interactions
 may also  be  needed to pin down the corrected e.v.
The numerical values of quark masses and meson masses needed to calculate
the entries of Tables (\ref{table:obsv}) and (\ref{table:obsv-shifts}) are displayed in 
Table (\ref{table:obsv-limits}).

\begin{table}[ht]
\caption{
\small Numerical results for some observables of the pion and the kaon. 
Where it has not been indicated, only the more complete 
set  [M3] was considered. 
The estimation of the strangeness content of the pion
is Pr.$s$-content $\pi$ .
e.v. refers to experimental or expected values.
The $\pi^0-\eta$ mixing angle ($\epsilon_2$) is calculated
in two cases 
 (I)  with $\theta_{ps}^{08} \equiv \theta^{ps}= 15^{\circ}$ 
and  (II) with $\theta_{ps}^{08}$ from eq. (\ref{thetaps08})
(its e.v. is the average 
value ($^*$) from the references in the line below).
Values (e.v.) of quark condensates  from
Refs.  \cite{jamin,flag,davies-etal,harnett-etal},
values for $\theta_{ps}$ from
 \cite{PDG}
and for $\epsilon_2$  $^*$
from \cite{feldman-etal-kroll,pi0-eta-etap,mixing-exp-1,tippens-etal,abdel-etal}.
} 
\centering  
\begin{tabular}{|  c || c  | c  |  c  | c ||  
  c  | c  |  c  | c ||
c | }  
\hline\hline 
Observable & $S_2$   &  $S_5$  
 &  $S_6$ & $S$,$G_0$ & 
  $V_2$   & $V_5$ & $V_6$  & $V$,$G_0$ &   e.v.
\\
\hline
\hline
 $M_{\pi^0}$(MeV)  [M2] &   135.0 & 135.3  & 135.1 & 136.4
& 
 135.5   & 136.1 &  135.6 &  137.1 & 
135   \cite{PDG}
\\
 $M_{\pi^0}$(MeV)   [M3]  & 133.5 & 134.2 & 133.7 & 136.4
&
  134.15 & 134.9  &  134.4 & 137.1  & 
\\
$M_{\pi^\pm}$ (MeV)  [M2]  & 135.2 & 135.4 & 135.3  & 
 136.7
&  
 135.7 &  136.2  &  135.8 &  137.4 & 
 \\
 $M_{\pi^\pm}$ (MeV)   [M3]  & 133.7  & 134.4 & 133.9
&  136.7 &
 134.4  & 135.0 &  134.5 &  137.4 & 
\\
 $M_{\pi^\pm}-M_{\pi^0}$ (MeV)  [M3]  & 
0.2 &  0.2 & 0.2 &  0.3
&   0.1 
& 0.1 & 0.1  &  0.3 &  
0.1 
 \cite{PDG,gasser-leutwyler-82,gasser-leutwyler-1985}
\\
\hline
  $M_{K^0}$ (MeV)   [M3]  & 498.5 & 498.5 & 498.5 &  499  & 
 498
   &  498 &   499 & 498 & 
498  \cite{PDG}
\\
 $M_{K^\pm}$ (MeV)  [M3]  & 490  & 491  & 493 &  490 
 &  486  &  487 &  488 & 490 & 
494  \cite{PDG}
\\
\hline
$F_\pi$ (MeV) & 99 &  99 & 99 & 102
& 
  100 &  101  &   101   & 103  &  92 
\\
$F_K$ (MeV) & 111  & 111 & 111 &  112 &   
 112  &  112 & 111 & 113 & 
 111  
\\
$(-<\bar{u}u>)^{1/3}$ MeV & 331  & 334  &  332 & 343
  &  336  &  338 & 336  & 347  & 
240-260
\\ 
$(-<\bar{d}d>)^{1/3}$ MeV & 333  & 335  & 333  & 344 
 & 338   & 339  & 338  & 349  & 
 240-260  
\\
$(-<\bar{s}s>)^{1/3}$ MeV$$ & 348  &  353 & 349  & 366
   &  352 & 356 & 353  &  369  & 290-300  
\\
$G_{qq\pi}$ &  3.3   & 3.2 & 3.2   &  3.4  &   
 3.3 &  3.3  & 3.3  & 3.6  & 
 \\
$G_{qqK}$ &  3.8   & 3.8    &  3.8  & 4.2 &
 3.9 & 4.0  & 3.9  & 4.3 & 
\\
\hline
$\theta_{ps}$ & -3.7 & -2.7  &  -3.6  & 0.0 &  
  - 3.4 & - 2.6  & - 3.4 & 0.0 & 
 (-11$^\circ$)-(-24$^\circ$) 
\\
$\epsilon_2$  (I)  & -0.8     &  -0.6  &  -0.8  & - & -0.8    & -0.8  &  -0.8 & - &  $(-1^\circ)^*$

\\
$\epsilon_2$  (II)  & -3.3    &  -3.2 & -3.3 &  - &  -3.6  &  -4.6 & -3.4  & - &
\\
\hline
Pr.$s$-content $\pi$ & 
$6\%$ & $10\%$ & $10\%$  &  0 &      $16\%$ &     $16\%$ &   $16\%$  &  0 &  
\\
\hline
$\chi^2_{red}$ (with $<\bar{q}q>$) & 103  & 110  & 103  & 146    &
122  & 129 &  123  & 164 &
\\
\hline
$\chi^2_{red}$ (without $<\bar{q}q>$) & 29    & 30   & 27  & 
51 
& 37    &  41  & 39 & 56 & 
\\
\hline \hline
\end{tabular}
\label{table:obsv} 
\end{table}
\FloatBarrier

{

In Figure (\ref{fig:Fps})
the charged pion decay   constant, $F_\pi$, is presented  
a function of 
 the strange quark effective mass 
 for the three sets $S_2, S_5$ and $S_6$ for the case of $M3$
defined in eq. (\ref{M2M3}).
It is noticed
a  clear decrease of the pion  decay constant with an increase 
of the strange quark (effective) mass.
Some few  results obtained from ChPT, however,
indicate that the pion decay constant should actually increase with increasing
strange quark mass $m_s$ or 
kaon mass $M_K^2$ \cite{guo-lutz,bijnens-fpi}.
These available results  from ChPT  were obtained 
with quite  large uncertainties in the 
knowledge of some lec's, $l_4, l_5$ and $l_6$
and a more complete investigation about this issue is missing.
There is not extensive specific results from lattice QCD that disentangle  fully
the dependence on the strange quark mass from other variables such as the pion mass.

\begin{figure}[ht!]
\centering
\includegraphics[width=120mm]{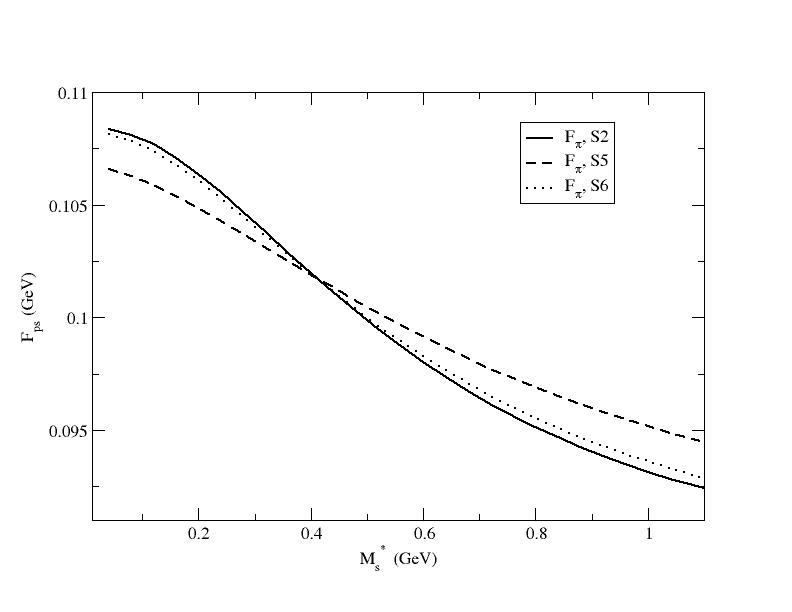}
 \caption{ 
\small
$F_\pi$ 
 a function of $M_s^*$, arbitrarily varied, for the same sets of parameters of the 
previous figures, $S_2, S_5$ and $S_6$.
All the other parameters $M_u^*,M_d^*$
and coupling constants
 are obtained self consistently.
}
\label{fig:Fps}
\end{figure}
\FloatBarrier

}

\subsection{ Strangeness contribution for the masses of  the u, d constituent
 quarks and  pion}

In the  Table \ref{table:obsv-limits}
further numerical results obtained from the change in $M_s^*$
 are displayed.
In the upper part of the Table there are 
particular values for  the effective quark effective masses 
$M^*_u, M^*_d$ calculated self consistently (for M2 and M3). 
The values obtained from 
complete self consistent calculation -  that   reproduces the neutral and charged
pion and kaon masses -  are identified by $G2$.
Both calculations for $M2$ and $M3$
correspond
to different ways of taking into account $G_{88}$ as responsible 
for a strange-up/down asymmetry.
The resulting up and down quark effective masses 
when $M^*_s \to 0$,  $M_s^* \to m_{s}$ and $M_s \to \infty$
are also shown.
Note that the self consistent calculation of $M^*_{ch.L.}$ ,
that is a flavor symmetric  limit with degenerate quark masses,
for the chiral limit,
$m_u=m_d=m_s=0$,
provides a lower value for the quark effective masses
than the value for $M_f^*(M_s^*\to 0)$ (f=u and d).
 Note also  that self consistent calculation
for $M_{u,d}^*$  can easily provide values lower than
their value in the  chiral limit.
 Values of the  neutral pion masses 
are shown in the same limits of  $M2$ and $M3$ 
- self consistent results  identified by $(G2)$ - 
and also 
for $M_s^* = 0$, $M_s^*=m_s$ and $M_s^* \to \infty$.
The analysis is basically the same as that for the up and down constituent 
or dressed quarks above, being however 
that  in the chiral limit  $M_{\pi}^{ch.L} = 0$ since the pion is a Goldstone boson. 
The behavior of the charged pion mass  is basically 
the same as the neutral pion mass
as shown above.

\begin{table}[ht]
\caption{
\small 
Numerical results for up and down quark effective masses
and for the neutral pion mass,
several of them obtained by varying freely the strange quark effective mass,
$M_s^*$,   and some of them obtained fully self consistently, identified by $G2$:
${M_f^*}$ and $M_{\pi_0}$ for $(M2)$ and $(M3)$
and $M^*$(ch.lim.).
Value for (e.v.) for $M_\pi^0$ from Refs.   \cite{PDG,gasser-leutwyler-82}.
} 
\centering 
\begin{tabular}{| c || c  | c  |  c  | c ||  
  c  | c  |  c  | c ||
c |}  
\hline\hline  
Observable/M3  & $S_2$   &  $S_5$  
 &  $S_6$ & $S$,$G_0$ & 
   $V_2$  & $V_5$ & $V_6$  & $V$,$G_0$ &   e.v.
\\
\hline
\hline
  ${M_u^*}$ M3-(G2)  (MeV) & 367  & 377 & 368 &   405  &  
 385 &  393 & 387 & 422 & 
\\ 
${M_u^*}$ M2-(G2)   (MeV) & 386   & 394 & 387  &  405 &   
 401  &  406 & 402 &  422 & 
\\
${M_u^*} ({M^*}_s \to 0)$ MeV & 618   & 512   & 595  &  405 
& 651 & 537  &  626 & 422 & 
\\
${M_u^*} ({M^*}_s \to m_{0,s})$ MeV & 563  &  491 & 547  &  405
 & 579  &  508  & 563  & 422 & 
\\
${M_u^*} ({M^*}_s \to \infty)$ MeV &  290  &  310 & 295  &
 405  & 300  & 316 & 304  & 422  & 
\\
\hline
 ${M^*_d}$  M3-(G2)   (MeV) & 375 & 384 & 378  & 415 &  
 394 & 402  & 395  & 431  &  
\\
 ${M^*_d}$ M2-(G2)  (MeV) & 396   & 399 & 396  &  415 &  
410  &  415  & 411  & 431 & 
\\
${M_d^*} ({M^*}_s \to 0)$ MeV &  625  & 520 &  602  & 415
 & 657 &  544 &  632 & 431 & 
\\
${M_d^*} ({M^*}_s \to m_{0,s})$ MeV & 555  &  491  &  541  &  415
& 585  &  514 & 568  & 431 & 
\\
${M_d^*} ({M^*}_s \to \infty)$ MeV &  305 & 320  & 305 &  415
 & 314  & 328  &  316 & 431 & 
\\
\hline
 ${M_s^*}$  M3-(G2)  (MeV)  &   555  &  567   &  558  &  612 & 
 566  &  581 & 569 & 625  & 
\\
  ${M_s^*}$  M2-(G2)  (MeV)  & 560 & 570 & 563  & 612 & 
  600 & 595  &   604 &  625& 
\\
\hline
ch.lim. $M^*_{ch.L.}$ (MeV) & 381 & 381 & 381 & 381
  &  415  &  415 &  415  &  415 &
\\
\hline \hline
 $M_{\pi^0}$ M3-(G2)   (MeV) & 133.5 & 134.2 & 133.7 & 136.7
&
  134.2 & 134.9  &  134.4 & 137.4  &  0.135 
\\
 $M_{\pi^0}$  M2-(G2)  (MeV) &   135.0 & 135.3  & 135.1 & 136.7
& 
 135.5   & 136.1 &  135.6 &  137.4 & 
\\
$M_{\pi^0} (M^*_s \to 0)$ MeV & 158  & 147  & 156  & 136.7  &
162  & 149  & 159 &  137.4 & 
\\
$M_{\pi^0} (M^*_s \to m_{0,s})$ MeV &  150  &  144 &  149 &  136.7
 & 150  & 144 & 149 &  137.4 & 
\\
$M_{\pi^0} (M^*_s \to \infty)$ MeV & 129   &  129 &  129 &  136.7 & 
129  & 129 & 129 & 137.4  & 
\\
\hline \hline
\end{tabular}
\label{table:obsv-limits} 
\end{table}
\FloatBarrier

Different ways of defining strangeness  and flavor asymmetry content of the 
constituent, or dressed, up  and down quarks and of the pion can be envisaged.
These mass differences  discussed below are different from the usual 
strange-sigma terms, either for the constituent quarks u and d
(as responsible for nearly 1/3 of the nucleon mass)
and for the pion. 
 It  becomes useful to define particular differences 
of values that might correspond to variations with 
specific  meanings.
From 
here on, these quantities will  be referred as to $T_{ff} =  M_f^*, M_{\pi}$
and also  $G_{ff}$.
Furthermore these quantities defined below can also apply to
the coupling constants  $G_{ff} $ that present basically  the same 
behavior of the up, down quark effective masses when varying 
$M_s^*$.
We will make use of the following 
 differences of a quantity $T_{ff}$ 
to characterize a specific
s-content of the  up and down quarks and of  the pion:  
\begin{eqnarray} \label{s-M2-M3} 
  \Delta_{s}^{2,3} &=& T_{ff} (M3) -  T_{ff} (M2) ,
\\
 \label{s-M0}
 \Delta_{s}^{0} &=& T_{ff} ({M^*_s}') - T_{ff} (M^*_s=0), 
\\
\Delta_{s}^{m_0} &=&
 T_{ff} ({M^*_s}')
 -  T_{ff} (M^*_s=m_s),
\\
 \label{s-Minf} 
\Delta_s^{\infty} &=& T_{ff} ({M^*}_s') - T_{ff} ({M^*}_s \to \infty) .
\end{eqnarray}

These mass differences will be exhibited in Table (\ref{table:obsv-shifts}).
First, note that
the difference between the two curves,
 $M2$ and $M3$, 
can be considered as a 
first measure of the effect of the flavor- asymmetry (for 
the strange quark) for  sea quarks
in the coupling constants
due to coupling $G_{88}$.
This 
quantity $ \Delta_{s}^{2,3}$  
 is  defined 
at the physical point.

Deviations  with respect to the limit 
 in which the strange quark effective  mass is zero  
is encoded in $\Delta_{s}^{0}$, and it 
could be interpreted as the overall contribution of 
the strange quark effective mass to the observable $T_{ff}$.
The deviation of the quantity $T_{ff}$ with respect to the point where 
strange quark condensate goes to zero
was   defined  as
  $\Delta_{s}^{m_0}$.
This is an effective measure of the strange quark condensate
in the constituent/dressed quark or pion.
These mass differences, however,
 are not fully  extracted in  physical points in the 
sense that in these limits, $M^*_s=0$ and $M^*_s=m_s$, 
 there are no non trivial
solutions for the full self consistent problem 
and kaons are not bound.
The mass difference $\Delta_s^{\infty}$ provides 
a dynamical way 
 to measure the shift on 
the value of $T_{ff}$ due to the the strange quark effective mass.
In its definition one considers  the limit in which 
the strange quark effective mass  goes to infinite $T_{ff} (M^*_s \to \infty)$.
Curiously this is a well definite limit with interesting physical appeal,
since in this limit the strange quark degrees of freedom should be frozen.

The same reasoning done above  for mass differences, with  $T_{ff}$,
applies for the  $G_{uu}$ and $G_{ss}$
because these coupling constants present a very similar behavior
with the change in  $M_s^*$.

\subsection{
 Strangeness content of pion mass and the
 $\pi^0 - \eta$ mixing}

To analyze further
the mixing (\ref{mix-rot-p0-p3})
in the physical pion state $|\pi^0>$,
let us define the states:
\begin{eqnarray}
|P_3> &=& \frac{1}{\sqrt{2}}
( | \bar{u} u > - | \bar{d} d > ),
\\
|P_0> &=& \frac{1}{\sqrt{3}}
( | \bar{u} u > + | \bar{d} d > + | \bar{s} s >),
\\
|P_8> &=& \frac{1}{\sqrt{6}}
( | \bar{u} u > + | \bar{d} d >  - 2  | \bar{s} s >).
\end{eqnarray}
With the mixing (\ref{mix-rot-p0-p3})
the physical pion state, for the $\pi^0-\eta$ mixing, can be written as:
\begin{eqnarray} \label{pi0-mix}
| \pi^0 >  &=& \frac{1}{\sqrt{2} }
 \left[ 1 + a \right]  |\bar{u} u > 
-
\frac{1}{\sqrt{2} } \left[ 1 - a \right]
|\bar{d} d> 
- \sqrt{2} a | \bar{s} s > ,
\nonumber
\\
a   
&=& \frac{1}{3} [ \epsilon_1 + \sqrt{3} \epsilon_2 \cos(\theta_{ps}) ],
\nonumber
\end{eqnarray}
The sea quark correction to the neutral pion
 state $|\bar{d}d>$ has  the opposite  sign of the one for the
state $|\bar{u}u>$.

As stated above, in phenomenology $\epsilon_1$ (or equivalently 
$\epsilon'$) is considerably smaller than $\epsilon_2$ and it will be neglected.
Let us assume these states for a constituent quark model, that must be 
further specified,
provide masses by applying  energy operator in a rest frame.
For orthonormal quark-antiquark singlet states, it may be considered to yield:
$<\bar{q} q |\hat{H} |\bar{q} q > \simeq 2 M_q $.
The pion seems to be a particle for which the CQM does not provide
 good results given its very low mass in the hadron spectrum.
What happens to the u-d sector with the valence quarks is not really important
for this estimation of the $\pi^0-\eta$ mixing.
 In what concerns a possible strange quark content for the pion
in eqs. (\ref{pi0-mix}) a usual value for the strange quark mass  
can be assumed, $M_s^*\simeq 450$ MeV.
 \footnote{ It can be seen, however, in Table (\ref{table:obsv-limits}) that the 
quark-mixings (up, down and strange) can  lower the quark effective masses.
This also leads to additional strangeness content of up and down (constituent) quarks.
So, one could even ask whether all these possible  mixings, for some reason
in the case of the 
pion,  make the pion mass 
considerably smaller than the other pseudoscalar mesons
 down to nearly $135$ MeV.
The answer seems to be, of course,  no,
although one may ask the opposite question, i.e. why not? (why the 
neutral pion does not mix so strongly with the other neutral pseudoscalar).
The neutral pion makes part of a iso-triplet state that for some 
reason (mainly isospin symmetry) may
protect the neutral pion from strong mixings that do  not take place for the charge pions.
However these types of questions
 will not be addressed further in the 
present work.
Goldstone boson  masses are rather guided by the GellMann Oakes Renner
relation.
}
  These valence and sea quarks, however,  do not  really need to be 
 quasi-particles since they are all confined.
In this case the strange sea quark   contribution for the 
pion mass must be proportional to 
\begin{eqnarray} \label{delmpi-s1}
\Delta_\eta m_{\pi^0} \simeq   4  a^2 M_s.
\end{eqnarray}
 It can be written that the shift in the up or down constituent quark masses due to the 
mixing with the strange is of the order of
\begin{eqnarray}
\Delta_\eta  M^*_{u,d} \sim \frac{3}{4} \Delta_\eta m_{\pi^0}.
\end{eqnarray}

\subsection{ Mass differences: Numerical Results}

Some mass differences for the 
  up  and down quarks,
 calculated according to eqs. 
(\ref{s-M2-M3}-\ref{s-Minf}), are exhibited in  Table (\ref{table:obsv-shifts}).
 Two different sources of 
 changes in the   up and down quark  effective  masses 
can be immediately identified.
Firstly  the shift in effective masses due to the 
explicit chiral symmetry breaking by mean of the quark 
current masses, and secondly 
the shift in the quark effective masses due to the 
flavor-dependent coupling constants by means of the 
self consistent procedure.
By denoting the quark effective mass in the chiral limit ($m_0=0$)
- that are degenerate - 
by $M_{ch.L.}$
these two mass shifts can be written respectively as:
\begin{eqnarray} \label{DelMm0}
\Delta_{chL}^{(f)} &\simeq& M_f  - M_{ch.L.},     
\\ \label{DelMM}
\Delta_{G_{ij}}^{(f)}  &\simeq&  M^*_f  - M_f,
\end{eqnarray}
where $M_f$ are solutions for the first gap equation $G1$ and     
$M^*_f$ are solutions for the second gap equations $G2$. 
This second quantity, $\Delta_{G_{ij}}^f$
for the up and down quarks, is mainly due to strangeness content 
of the coupling constant and, less importantly, also 
correspond to providing a u-content (d-content)  for the d (u) 
quark effective mass.
 The numerical values for the 
neutral  pion mass differences, eqs.  (\ref{s-M2-M3}-\ref{s-Minf}),
are also shown in Table (\ref{table:obsv-shifts}), and they
present the same relative  behavior of the mass differences of 
up and down quarks.
The mass differences are sizable are independent
of the mixing interactions $G_{i\neq j}$, that should, by the way,
contribute as well. This complete calculation is not performed in the 
present work.
It  is interesting to compare these 
mass differences  with results
for 
the strangeness content of the pion of eqs. (\ref{delmpi-s1}) 
due to 
the mixing with the $\eta$ via interaction $G_{38}$.
The mass shift $\Delta_\eta m_{\pi^0}$ is presented for 
two different pseudoscalar angle mixing $\theta_{ps}$ (the 
one for the $\eta-\eta'$ mixing).
(I) for $\theta_{ps}=15^{\circ}$ \cite{PDG}
and (II) for $\theta_{ps}$ shown in Table (\ref{table:obsv})  obtained in the calculations.
 
 {
The comparison with results for the strangeness content of the pion 
 from other works that consider different frameworks may not be
direct. 
In the following, we summarize some results found in the literature for the
strangeness content of the pion.  
As discussed above, the contribution of the kaon cloud for the pion mass was found 
to be negligibly small, of the order of $1$ MeV in Ref. \cite{cloet-roberts}.
Sigma terms have been calculated in different approaches
from NJL, constituent quark models, Chiral Perturbation theory
 and more recently lattice QCD
for example in:
\cite{NJL-sterm,gunnar-etal-latt,strangeq-latt,pion-sigma,chpt,sigma-sainio,bijnens-dhonte}.
Specifically for the pion strange sigma term, $\sigma_s^\pi$,
lattice QCD calculations \cite{gunnar-etal-latt} provided 
small values with large uncertainties being even compatible with zero.
The pion-strange sigma term for the  flavor dependent NJL model 
however is not exhibited in the present work
because the normalization considered, eq. (\ref{Gnorm-ij}),
leads to different interpretations and quite ambiguous results.

In ChPT and ChPT with unitarization there are two types of calculation.
Firstly, the s-sigma term  has been investigated as the zero momentum limit 
of the strange scalar form factor of the pion,
up to $p^{(6)}$ order,   to be of the order of few percent of the pion mass
\cite{chpt,sigma-sainio,bijnens-dhonte}.
There are imprecisions in the contributions of the  fourth and sixth
order contributions because of the 
 imprecision in the knowledge of lecs $(l_4,l_5, l_6)$.
It is reasonably
 comparable to $\Delta^0_s$, $\Delta_s^{2,3}$ or $\Delta_s^{\infty}$
 in Table (\ref{table:obsv-shifts}).
Secondly, the strange quark contribution for the pion mass was 
calculated in ChPT in \cite{rkaiser1,rkaiser2} up to $p^{(5)}$ order.
 It is not exactly the s-sigma term but it must contain part of the 
s-sigma term content.
The resulting contribution  can be written
in a similar shape of the result found in the present work as:
\begin{eqnarray} \label{mpi2-kaiser}
M_\pi^2  = ( 1 +  \Delta_s) m_\pi^2,
\end{eqnarray}
being $\Delta_s \sim  0.14 \to 0.31$.
From this, the correction for the pion mass can be written as:
$M_\pi =
\simeq  m_\pi + \delta_s$.
This contribution can be comparable to
the mass-difference $\Delta_s^0$ shown in the Table.
To summarize these results the following values can be 
considered:
\begin{eqnarray}
LQCD \;\; & \sigma_s^\pi &\;\;\;\;  6 (33) \; MeV  \;\;\; \mbox{at}\; m_\pi = 149.7MeV\;\;  \cite{gunnar-etal-latt}
\\
ChPT \;\; & F_{Ss}^{\pi}(t=0) = \sigma_s^\pi & \;\;\;\; 0 - 12 MeV \;\;\;\; \cite{chpt,sigma-sainio,bijnens-dhonte}
\\
ChPT  \;\; & \delta_s &  \;\;\;\;
9  - 19  MeV\;\;\;\; eq. (\ref{mpi2-kaiser}) \;\;\;\; \cite{rkaiser2}
\end{eqnarray}

}

\begin{table}[ht]
\caption{
\small Numerical results for up and down quark effective mass differences 
and neutral pion mass differences defined
in  eqs. (\ref{s-M2-M3}-
 \ref{DelMM}).
In the last two lines the strange quark mass contribution for the pion mass
by means of eq.  (eq.(\ref{delmpi-s1}))
for two different mixing angles $\theta_{ps}$:
(I) for $\theta_{ps}=15^{\circ}$ \cite{PDG}
and (II) for $\theta_{ps}$ shown in Table (\ref{table:obsv})  obtained in the calculations.
 } 
\centering 
\begin{tabular}{| c || c  | c  |  c  | c ||  
  c  | c  |  c  | c |
 }  
\hline  
Observable [M3]  & S2   &  S5  
 &  S6 & S-$G_0$ & 
   $V$-2   & $V$-5 & $V$-6  & $V$-G0 
\\
\hline \hline
$ \Delta_{chL}^{(u)}$ \; (MeV) &  24  &  24 &   24 &     24
  & 7 & 7 & 7  & 7 
\\
$\Delta_{chL}^{(d)}$ \; (MeV) & 34  &  34 &  34 & 34 
& 16  & 16  & 16   & 16 
\\  
$\Delta_{G_{ij}}^{(u)}$  \; (MeV)
 & - 38   &  -28 & -37 & 0
& -37  & - 29 & -35 & 0 
\\
$\Delta_{G_{ij}}^{(d)}$  \; (MeV)
 & -40   & -31  & -37 & 0
& -37 & -29 & -36 &  0 
\\
\hline \hline
$\Delta_s^{2,3} ( {M_u^*})$     \;  (MeV)  
    & -19 
& -17    &  -19 & 0 & 
-16    & -13  & -15   &  0  
\\  
$\Delta_s^0 ( {M_u^*})$  \;  (MeV)  
    &  -251  & -135  &  -227  &  0
  &  -266  & - 144  &  -239  & 0 
\\  
$\Delta_{s}^{m_0} ({M_u^*})$   \;  (MeV)   & -194  &-114   & -179 & 0  &
 -194  & -115  & -176 & 0 
\\ 
$\Delta_s^{\infty} ({M_u^*})$    \;   (MeV)   &  77  &  67  &  73  & 0  &
85    & 77  & 83 & 0 
\\
\hline\hline
 $\Delta_s^{2,3} (M_{\pi}^*)$  \;  (MeV)  & -1.5  & -1.1  &  -1.4 &  0  &
 -1.3   & -1.2  & -1.2 & 0 
\\ 
$\Delta_s^0 ( M_{\pi}^*)$  \;  (MeV)   & -24   & -13   & -22  & 0 
&
-28   &  -14  & -25 & 0
 \\  
$\Delta_{s}^{m_0} (M_{\pi}^*)$  \;  (MeV)   &  -16  &  -10 & -16  & 0  &
-16    &  -9 & -14 & 0 
\\  
$\Delta_s^{\infty}  (M_{\pi}^*)$  \;   (MeV)   & 6 & 5  & 4  &  0 &
 5    &   6   &  5 & 0 
\\
\hline\hline
$\Delta_\eta m_{\pi^0}$   (I) (MeV)
   & 0.11 & 0.10  & 0.11  & 0   & 0.11 &  0.11 &  0.11 & 0
\\
$\Delta_\eta m_{\pi^0}$  (II) (MeV)
  & 1.9   & 1.8 & 2.4  & 0   & 2.4   &  3.7  &  2.1 & 0
\\
\hline
\end{tabular}   
\label{table:obsv-shifts} 
\end{table}
\FloatBarrier

\section{ Summary}

A self consistent calculation for the quark effective masses 
and flavor-dependent coupling constants of the NJL model 
 was employed to investigate
the role of  
  the strange quark effective  mass on the up and down constituent quarks 
and on the pion masses.
 The fully self consistent calculation
was done to fit the parameters of the model to reproduce neutral (or charged)
pion and kaon masses.
In this step, several observables were calculated to assess  the
reliability of the model  for different sets of parameters.
The strong value of the  coupling constant of reference,
 $G_0=10$ GeV$^{-2}$,
in one hand, helps with the convergence of the self consistent 
numerical calculation but, 
in the other hand, induces considerable large values of the 
quark-antiquark condensates. 
Flavor dependent coupling constants, nevertheless,
improve their values with respect to the ones obtained with $G_0$.
Quark effective masses get lower and they require a slightly  larger cutoff to 
make possible the description of light mesons masses.
{A further consequence of this larger value of the coupling constant of reference
$G_0$, associated to the use of 
the three-dimensional cutoff, is  the relatively low value of the cutoff
which limits the large UV  three-momenta instead of the four-momenta.
}

Because of  the quantum mixing
it was possible to estimate the s-content of the 
up and down constituent/dressed quarks
as well as of the  pion even in the absence of 
mixing-type interactions.
Different ways of extracting the strangeness contributions for the 
up and down quark  effective masses and pion masses
were analyzed in this work by means of mass differences. 
{Besides the resulting effect from the renormalization of the wave function $Z_\pi$,
several mass differences were exhibited.}
The mass difference $\Delta^{2,3}_s$  
 provides the change in the pion (or up and down effective) mass
when exchanging the value of   the interaction $G_{88}$,
 without any flavor-dependence or mixing effect (M2),
  $G_{88}=10$GeV$^{-2}$,
by a value that takes into account the strange sea quark dynamics,
(M3) with  $G_{88}=G_{88}(M^*_s)$.
The mass difference $\Delta_s^\infty$ 
corresponds  to a sort of
 dynamical criterium to define a strangeness contribution 
for  the pion
mass  (or 
up and down effective masses).
 The same analysis is valid for the flavor dependent coupling constants
$G_{ff}$.
It is important to note that 
 the mixing investigated in the present work is responsible for 
the strangeness in the up and down quark sector corresponds to 
a  different mechanism 
from 
 the quark mixing induced usually given by 
 the  CKM matrix or from the instanton-induced determinantal effective interaction.
These  mechanisms  should add to each other.
The whole  procedure does not necessarily  lead to a simple rotation 
of dressed quarks 
  but it  might  involve some other transformation,
 like a dilation.
As a whole, results may  overestimate the contribution of the 
strange condensate to the pion structure  as compared to 
available lattice QCD, the kaon cloud for
up and down constituent  quarks
 and ChPT expectations \cite{gunnar-etal-latt,cloet-roberts,cheng-1989,chpt,sigma-sainio,bijnens-dhonte}.
This can be attributed mostly 
to the normalization (\ref{Gnorm-ij}) adopted in this work that easily strengthen 
the flavor-symmetry breaking contributions from the polarization process.
The chosen normalization garantees, nevertheless,  the strength of the 
flavor-dependent component was assessed with 
respect to the value of reference $G_0$.

The mesons mixings, $\pi^0-\eta$ and $\eta-\eta'$, 
were the only observables for which the mixing type interactions $G_{i\neq j}$ were
considered. 
Angle mixings were calculated to 
reproduce mesons mass differences.
The $\eta-\pi^0$ mixing was  associated to 
a strange quark  content of the pion.
It is interesting to compare resulting  estimations  
with the  mass differences 
for 
the strangeness content of the pion mass.
 These values have comparable order of magnitude, 
as seen in Table
(\ref{table:obsv-shifts}).
 However the physical origins are somewhat different.
It is important to note that  in the estimations of  meson mixings, the  binding energies and 
(valence) quark kinetic energies were neglected.
It looks that 
 the strangeness contribution for the pion mass extracted from 
the quantum mixing contributions (Table (\ref{table:obsv-shifts})) might be associated to 
an upper bound because of the normalization adopted for the coupling constants
$G^n_{ij}$.
{
The chosen  normalization for the coupling constants, at the 
flavor symmetric point ($m_u=m_d=m_s$)
eq. (\ref{Gnorm-ij}), provides quite similar results to the normalization adopted previously
\cite{FLB-2021a} although it provides a faster convergence of the self-consistent calculation.
The charged pion  decay constant was also calculated
as a function of the strange quark effective mass.
The resulting behavior was found to be the opposite of that obtained from 
ChPT \cite{guo-lutz,bijnens-fpi}.
These results from ChPT are nevertheless preliminar 
because some of the lec's involved in these calculations are not 
really well determined. 
Therefore it is not clear to what extent this behavior of $F_\pi$ with increasing $M_s$
 is  a shortcoming of the present model
and further physical input is needed to correct this behavior as it occurs for 
NJL-calculations for 
heavy mesons decay constants.
}
The
mesons  mixings induced by the mixing interactions ($\Delta_\eta m_{\pi^0}$)
might be understood as a minimum value because it neglects further 
mixing effects due
 to the quantum mechanical mixing and, eventually, other interactions such as the 
 't Hooft interaction.
 Besides that, due to the self consistent character of the calculation,
results  contain   mixings of the type 
$U_{us}, U_{ds}$ and  $U_{ud}$.
 Although the
NJL-model does not exhibit   confinement of quark and gluons,
 it provides an interesting and quite appropriate  effective way of 
investigating aspects of  hadron structure and dynamics.
It is   not clear whether or how confinement would imply  modification(s)
in the different mixings interactions and mechanisms.
A complete account of the flavor dependent - NJL model with 
all  mixing type interactions, $G_{i\neq j}$ and $G_{f_1\neq f_2}$,
will be reported in another work.

\section*{Acknowledgements}

F.L.B.  thanks short discussions
 with 
 C.D. Roberts,  B.O. El Bennich and J. Bijnens.
The author 
  is member of
INCT-FNA,  Proc. 464898/2014-5
and  he acknowledges partial support from 
CNPq-312072/2018-0 
and  
CNPq-421480/2018-1.

\end{document}